\documentclass[journal,twoside,web]{ieeecolor} 

\usepackage{generic}
\usepackage{cite}
\usepackage{amsmath,amssymb,amsfonts}
\usepackage{algorithmic}
\usepackage{graphicx}
\usepackage{algorithm,algorithmic}
\usepackage{hyperref}
\hypersetup{hidelinks=true}
\usepackage{textcomp}
\usepackage{orcidlink} 

\def\BibTeX{{\rm B\kern-.05em{\sc i\kern-.025em b}\kern-.08em
    T\kern-.1667em\lower.7ex\hbox{E}\kern-.125emX}}
    
\usepackage{comment}
\usepackage{siunitx}
\DeclareSIUnit{\belmilliwatt}{Bm}
\DeclareSIUnit{\dBm}{\deci\belmilliwatt} 

\begin{document}
\title{Bidirectional UWB Localization: {\huge A Review on an Elastic Positioning Scheme for GNSS-deprived Zones}}

\author{Cung Lian Sang\orcidlink{0000-0002-6144-9544}, Michael Adams\orcidlink{0000-0002-4986-3654}, Marc Hesse\orcidlink{0000-0002-9500-3284}, \IEEEmembership{Member, IEEE,} and Ulrich R\"uckert, \IEEEmembership{Member, IEEE} 
\thanks{We acknowledge support for the publication costs by the Open Access Publication Fund of Bielefeld University and the Deutsche Forschungsgemeinschaft (DFG).}
\thanks{The authors are with the Cognitronics and Sensor Systems Group, CITEC, Bielefeld University, Inspiration 1, 33619 Bielefeld, Germany (e-mail: csang@techfak.uni-bielefeld.de; madams@techfak.uni-bielefeld.de; mhesse@techfak.uni-bielefeld.de; rueckert@techfak.uni-bielefeld.de). (Corresponding author: Cung Lian Sang)}
\thanks{\textit{\textbf{This article has been accepted for publication in IEEE Journal of Indoor and Seamless Positioning and Navigation. This is the author's accepted version which has not been fully edited. Citation information: DOI 10.1109/JISPIN.2023.3337055}}}
}

\maketitle

\begin{abstract}
A bidirectional Ultra-Wideband (UWB) localization scheme is one of the three widely adopted design integration processes commonly used in time-based UWB positioning systems. The key property of bidirectional UWB localization is its ability to serve both navigation and tracking tasks within a single localization scheme on demand. Traditionally, navigation and tracking in wireless localization systems were treated as separate entities due to distinct applicable use-cases and methodological needs in each implementation process. Therefore, the ability to flexibly or elastically combine two unique positioning perspectives (navigation and tracking) within a single scheme can be regarded as a paradigm shift in the way location-based services are conventionally observed. This article reviews the mentioned bidirectional UWB localization from the perspective of a flexible and versatile positioning topology and highlights its potential in the field. In this regard, the article comprehensively describes the complete system model of the bidirectional UWB localization scheme using modular processes. It also discusses the demonstrative evaluation of two system integration processes and conducts a SWOT (Strengths, Weaknesses, Opportunities, and Threats) analysis of the scheme. Furthermore, the prospect of the presented bidirectional localization scheme for achieving precise location estimation in 5G/6G wireless mobile networks, as well as in Wi-Fi fine-time measurement-based positioning systems was briefly discussed.
\end{abstract}


\section{Introduction} \label{sec_introduction}

\IEEEPARstart{I}{n} outdoor environments, the Global Navigation Satellite System~(GNSS) is regarded as the de facto standard as well as the industry standard for positioning and navigation systems~\cite{Breeman2022Battles}. This is because GNSS is capable of providing the requirements of technical specifications and solutions for many sectors in governments, industry, and space agencies. However, the GNSS cannot provide sufficient and reliable location services in indoor and cluttered environments~(e.g. inside buildings, tunnels, underground, underwater, etc.)~\cite{Mendoza-Silva2019Meta}. The primary reason is that radio signal propagation in indoor environments is generally obstructed by obstacles such as buildings, walls, furniture, etc. Moreover, indoor environments are naturally loaded with closely located dense objects, which are movable including people as well as fixed ones. That causes multi-paths and signal blockage situations. As a consequence, it is infeasible to achieve the direct Line-of-Sight~(LOS) between the mobile device and the satellite base stations, which plays a central role in the GNSS location estimation process. Therefore, alternative technologies specifically designed for indoor or GNSS-denied environments/zones, which is commonly termed as Indoor Positioning Sytem~(IPS)~\cite{Mendoza-Silva2019Meta}, have been profoundly sought in recent years. 

Among the available technologies for IPS in the literature~\cite{Mautz2012Indoor, Mendoza-Silva2019Meta}, Ultra-Wideband (UWB) technology has been considered a promising and viable option for GNSS-denied environments~\cite{Alarifi2016Ultra}. The main reasons for this consideration include the UWB's ability to provide decimeter-level (i.e., centimeter-level) ranging accuracy, penetrate obstacles, support high data rates, potentially offer low-power and compact hardware, resist jamming, be immune to interference, and coexist with narrow bandwidth (NB) technologies~\cite{Yang2004Ultra, Win2009History}. Overview, time-based UWB positioning and navigation is a common practice in the system implementation process. In fact, there are essentially three topological perspectives on time-based UWB IPS~(Section~\ref{sec_SOTA_UWB_topos}). To clarify the concept of the mentioned three time-based UWB localization systems, we utilized the well-known GNSS system setup as a contrasting abstract reference in this article. In the GNSS system setup, in a nutshell, satellites periodically transmit signals, and the receiver on Earth estimates its location using those signals to navigate or reach the desired target. 
In consideration of this, the setup of UWB-based IPS, particularly for time-based localization systems, can be broadly categorized into three types~\cite{Ghavami2007Ultra, Sang2019Bidirectional}~(Section~\ref{sec_SOTA_UWB_topos}):
\begin{enumerate}
\renewcommand{\labelenumi}{\roman{enumi})}
    \item the GNSS-like UWB scheme, conventionally known as Downlink Time Difference of Arrival (DL-TDoA) \cite{Keating2019Overview} in classical mobile/cellular or satellite positioning systems,
    \item the inverted GNSS-like UWB scheme, also known as Uplink Time Difference of Arrival (UL-TDoA)~\cite{Keating2019Overview}, and
    \item the hybrid approach or bidirectional UWB scheme that flexibly integrates the two viewpoints into one topology, conventionally known as a Round Trip Time (RTT)~\cite{DeAngelis2016Positioning}.
\end{enumerate}

In brief, the GNSS-like UWB scheme (i.e., DL-TDoA) is commonly used in navigation systems, where a moving object is assisted or guided in reaching its destination by providing position information~\cite{Groves2013Principles}. In contrast, the inverted GNSS-like UWB scheme (i.e., UL-TDoA) is typically utilized in tracking systems, where mobile objects are observed and monitored through a central server or third party by extracting their location information. Consequently, the characteristics of navigation and tracking responsibilities within a positioning system are fundamentally distinct, often requiring separate implementation processes in practice. The bidirectional UWB scheme (i.e., RTT) flexibly combines these two perspectives (navigation vs. tracking) within a single system, making it interchangeable as needed for various applications.

The topologies of the UWB localization system regarding the GNSS-like scheme~(DL-TDoA)~\cite{Ledergerber2015Robot} and the inverted GNSS-like scheme~(UL-TDoA)~\cite{Zwirello2012UWB} have been extensively explored for industrial and academic use cases in recent years~\cite{Zandian2017Robot, Santoro2023UWB, Paetru22023FlexTDOA, Yang2022VULoc, Elsanhoury2022Precision, Hamer2018Self, Corbalan2019Chorus, Grosswindhager2019SnapLoc, Santoro2021Scale, Tiemann2016ATLAS, Tiemann2017Scalable, Tiemann2019ATLAS, Vecchia2019TALLA, Friedrich2021Accurate}. However, the bidirectional UWB localization scheme was typically mentioned as a classical method in the literature~\cite{Tiemann2016ATLAS, Zandian2017Robot, Hamer2018Self, Ledergerber2015Robot, Zwirello2012UWB}, without providing a specific implementation process, assuming that readers are already familiar with the concept. In fact, a thorough description of the bidirectional UWB system, from a theoretical standpoint to practical use cases, is generally lacking especially for the newcomers. Furthermore, the bidirectional UWB localization scheme was seldom considered a strategic localization scheme in the literature~\cite{Sang2019Bidirectional, Sang2022Dissertation}. Consequently, the potential of the bidirectional topology in the field has been excessively overshadowed. Indeed, the bidirectional scheme possesses several unique properties, applications, and prospects for use in diverse technologies beyond UWB (Section~\ref{sec_SOTA_UWB_topos} and \ref{sec_potential_beyond_uwb}). Therefore, this article reviews the aforementioned bidirectional UWB scheme from the perspective of an elastic positioning topology, and the scheme is dubbed a bidirectional UWB localization. It aims to introduce the fundamentals of UWB localization systems with a particular emphasis on bidirectional UWB localization, while also acknowledging the significance of its unidirectional counterparts. We attempt the content to be easily understandable and inclusive to a  broad audience interested in UWB, requiring minimum prior knowledge. Accordingly, this article offers a comprehensive review of bidirectional UWB localization, covering the foundational concept from the basics (Section~\ref{subsec_minimum_setup} and \ref{subsec_complementary_setup}) to its practical implication~(Section~\ref{subsec_digest_practice_impl}), using a modular design principle approach. Moreover, the demonstrative evaluation of the measurement results typically encountered in the practical deployment of UWB technology was also concisely described~(Section~\ref{subsec_exp_demo}). Likewise, the SWOT analysis of the scheme compared to its counterpart unidirectional approaches was also discussed in the article~(Section~\ref{subsec_SWOT_analysis}).

Based upon the perspective mentioned above, the article is organized as follows: Section~\ref{sec_SOTA_UWB_topos} explains the details of three topological schemes in the UWB positioning systems. It is, then, followed by the potential use-cases of the bidirectional scheme apart from the UWB technology in Section~\ref{sec_potential_beyond_uwb}. The system model of the bidirectional UWB localization using the modular design principle approach was addressed in detail in Section~\ref{sec_system_model} and the demonstrative evaluation results as well as a SWOT analysis of the scheme were discussed in Section~\ref{sec_eval_results}. Thereafter, the related work was addressed in Section~\ref{sec_related_work} and finally, the concluding remarks were drawn in Section~\ref{sec_conclusion}.

\section{Fundamental of UWB Localization Schemes} \label{sec_SOTA_UWB_topos}

In terms of the system integration process, localization systems based on radio wave technology~(i.e., positioning using UWB, GNSS, WiFi, Bluetooth Low Energy~(BLE), etc.)~can generally be grouped into two namely range-based vs range-free approaches. In a nutshell, the latter range-free methodology conventionally relies on algorithms such as fingerprinting or scene analysis techniques as location estimation methods~\cite{Alarifi2016Ultra, He2016WiFi}. In contrast, the former range-based positioning system first measures the range or distance. This is accomplished by multiplying a signal time-of-flight~(TOF) between a transmitter and a receiver with the speed of light (i.e., the constant electromagnetic wave traveling time in the air interface). Then, a dedicated positioning algorithm is applied to estimate the location of a concerned moving target based on the measured range. The UWB technology addressed in this article falls, in general, under the range-based approach. 

\subsection{Brief on Time-based UWB Positioning and Navigation} \label{subsec_time_based_loc}

The localization process of range-based positioning systems can be organized into two phases, namely: (i) the signal measurement phase (i.e., the ranging process mentioned above) and (ii) the location estimation phase (i.e., estimating a location of a moving target using positioning algorithm and the measured ranges in the former phase). Measurement in the former phase is commonly carried out using time of arrival~(ToA)~\cite{Sang2019Comparative, Shen2008Performance},  TDoA~\cite{Xu2006Position}, angle of arrival~(AOA)~\cite{Peng2006Angle, Zhao2021ULoc} or received signal strength~(RSS)~\cite{KimGeok2020Review, Mendoza-Silva2019Meta} techniques. Particularly for time-based UWB localization~\cite{Gezici2009Position}, the ToA and TDoA techniques are the classical choice for establishing the system implementation process. In contrast, the RSS approach is rarely used in UWB localization due to its poorer performance compared to the other techniques mentioned~\cite{Sang2019Bidirectional}. Likewise, the AoA method does not receive as good awareness as its counterpart ToA and TDoA approaches due to the necessity of a more complex system setup such as antenna arrays and beamforming techniques \cite{Yassin2017Recent, Zhao2021ULoc}.  

In terms of the location estimation phase, a wide range of algorithms have been examined in the literature. The most commonly used methods include Trilateration~\cite{Sang2019Comparative}, Multilateration~\cite{Sang2019Bidirectional, Norrdine2012Algebraic}, closed-form algebraic methods like least-square~\cite{Shen2008Performance}, Taylor-series~\cite{FOY1976Position}, and  Maximum likelihood~\cite{Leitinger2014Multipath}, as well as recursive and statistical positioning techniques like Extended Kalman Filter~(EKF)~\cite{Brown1972Integrated}, Unscented Kalman Filter~(UKF)~\cite{Gibbs2011Advanced} and particle filter~\cite{Gonzalez2009Mobile}. From the perspective of the system implementation process, there are primarily three topological schemes for the time-based UWB localization systems~(Section~\ref{subsec_three_topos}). In short, this article specifically addresses time-based UWB localization schemes extensively applied in industrial and academic projects.

\subsection{The Three Schemes of Time-based UWB Localization} \label{subsec_three_topos}

\begin{figure}[!t]
    \centering
    \includegraphics[width=8.8cm]{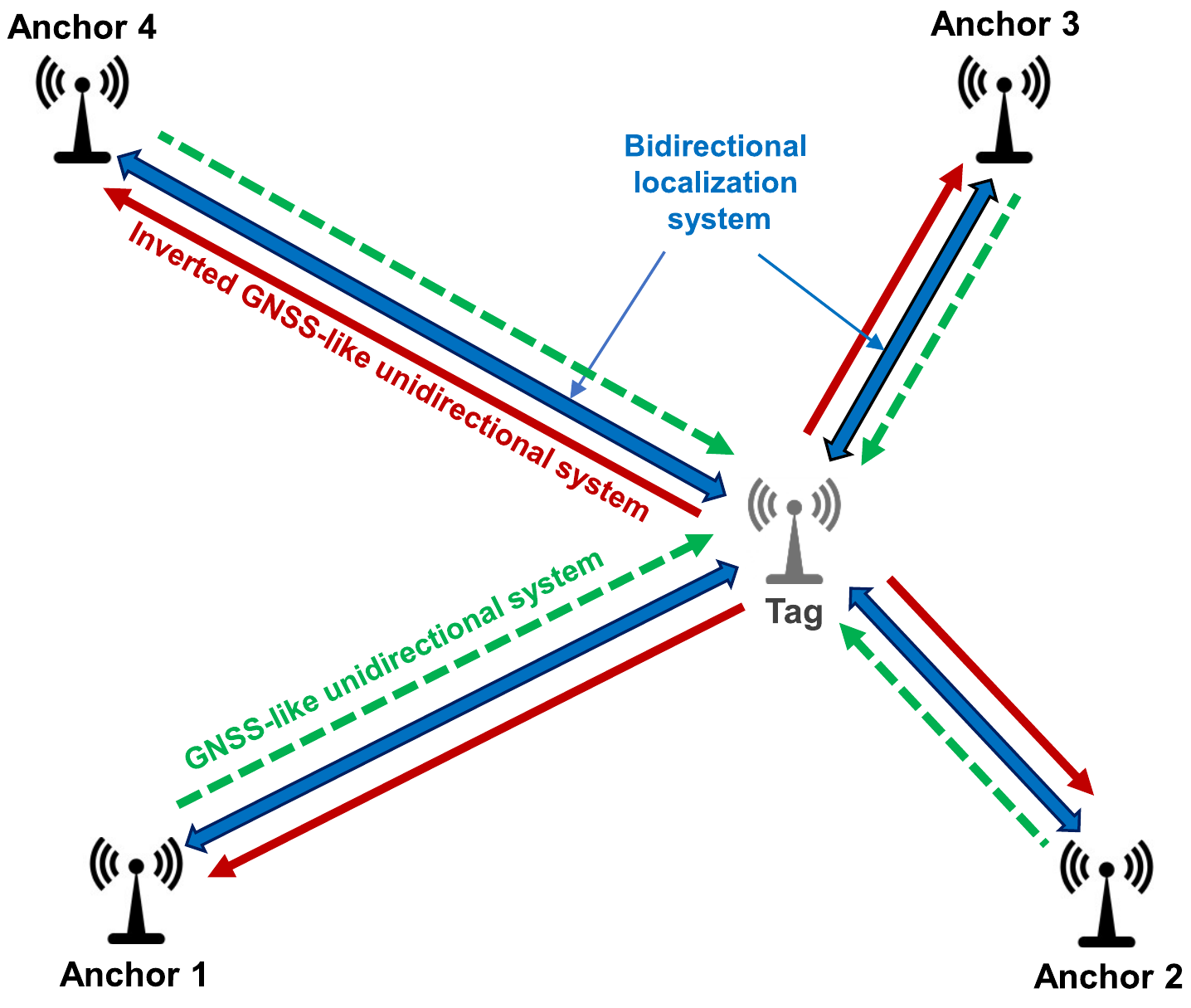}
    \vspace{-0.6cm}
    \caption{Simplified representation of three time-based topological schemes for UWB localization: (i) GNSS-like unidirectional UWB localization scheme~(GUULS), a.k.a DL-TDoA, (ii) inverted GNSS-like unidirectional UWB localization scheme~(IGUULS), a.k.a UL-TDoA, and (iii) bidirectional UWB localization scheme~(BULS). The arrows denote the direction of communication between UWB transceivers.}
    \label{fig_compare_three_uwb_schemes}
    \vspace{-0.5cm}
\end{figure}

There are mainly three schemes in the implementation process of a time-based UWB localization system as previously mentioned in Section~\ref{sec_introduction}. To the best of the authors' knowledge, the earliest definition of the stated three topological schemes specifically for UWB positioning and navigation was expressed in~\cite{Ghavami2007Ultra}. In the mentioned work, the aspect of the UWB localization system was defined as:~(i) a Network-based system (i.e.,~analogous to the GNSS-like approach or DL-TDoA), (ii) a Handset-based system~(i.e., akin to the inverted~GNSS-like method or UL-TDoA), and (iii) a Hybrid-system (i.e., the combination of the said two system integration processes namely GNSS-like and inverted GNSS-like systems). Following the analogous convention specified in~\cite{Ghavami2007Ultra}, this article categorizes the integration process of~ time-based UWB localization system into three topological schemes namely:~(i) GNSS-like unidirectional UWB localization scheme~(\textbf{GUULS}),~(ii) inverted GNSS-like unidirectional UWB localization scheme~(\textbf{IGUULS}), and~(iii) bidirectional UWB localization scheme~(\textbf{BULS}). For clarity, Fig.~\ref{fig_compare_three_uwb_schemes} shows the simplified diagrammatic illustration of the three UWB localization schemes mentioned in this article using a single non-stationary mobile device~(hereinafter referred to as tag) and its corresponding four known references or landmarks~(hereinafter referred to as anchors). In a nutshell, anchors are static fixed devices with their positions known to the system. In contrast, tags are remote devices commonly placed on mobile objects whose positions are of interest to determine by the localization system or the location-oriented service providers. Please note that this article focuses on active or device-based UWB localization schemes, which require attaching a tag to a mobile device or subject to determine its location. Device-free or passive UWB localization schemes~\cite{Jovanoska2013DeviceFree, Kilic2014DeviceFree, Santoro2023TagLess}, where a mobile subject's location can be estimated without a specific tag, are not within the scope of this article.

In a GNSS-like topology denoted as GUULS in this article, the operation of navigation within an environment can be performed by a tag device typically attached to a moving object. In GUULS~(or DL-TDoA), anchor nodes are placed at known fixed sites of areas, where the positions of moving remote objects are considered to be observed. Consequently, the anchors are responsible for periodically transmitting the UWB signals for the location estimation process in GUULS. The tag device then listens to the UWB signals sent from the anchors to estimate its current position to navigate to its targets. The term unidirectional in GUULS refers to one-way data communication between anchors and a tag (i.e., the senders are anchors and the receiver is a tag). Thus, GUULS is analogous to GNSS, where satellites correspond to transmit periodic radio signals in order to provide the location information of a mobile device on Earth to navigate the intended target location. The main use-case of GUULS in location service is the navigation or self-navigation purposes. The signal direction of data communication for UWB localization using GUULS is expressed as green dotted arrows in~Fig.~\ref{fig_compare_three_uwb_schemes}. Regarding the system integration process, the first GUULS in the context of UWB localization to the best of the authors' knowledge was described by the work in~\cite{Ledergerber2015Robot}. Subsequently, the scheme (i.e., GUULS or~DL-TDoA) has been extensively explored~\cite{Zandian2017Robot, Santoro2023UWB, Paetru22023FlexTDOA, Yang2022VULoc, Elsanhoury2022Precision, Hamer2018Self, Corbalan2019Chorus, Grosswindhager2019SnapLoc, Santoro2021Scale}. In the system integration of GUULS, the ranging phase (i.e. measuring the distances between anchors and tag) is conventionally based on the TDoA technique~\cite{Hamer2018Self} whereas the Multilateration method is normally applied for the location estimation process of the system~\cite{Zandian2019Ultra, Santoro2023UWB}. It should be noted that the mentioned ranging and positioning methods are not technically bounded, though a common practice in the field.

In contrast to GUULS, the inverted GNSS-like system integration process denoted as IGUULS in this article was widely deployed in the UWB-based localization systems. In IGUULS or UL-TDoA, tags (i.e., as opposed to anchors in GUULS) are responsible for transmitting UWB signals for the location estimation process. Then, the anchor nodes are accountable for computing the position of the remote moving object by using the sent UWB signals from the tag. Again, the term unidirectional in IGUULS refers to one-way communication between UWB transceivers~(i.e., the sender is a tag and the receivers are anchors in IGUULS) for location estimation purposes. The core use case of IGUULS in location service is a tracking scenario where the central server on the anchors' side is capable of simultaneously tracking the motion of all available tags within its system coverage area. The direction of the signal for data communication in UWB localization using IGUULS is depicted as red arrows in~Fig.~\ref{fig_compare_three_uwb_schemes}. In terms of the integration process for UWB-based localization, the first IGUULS was proposed and implemented in~\cite{Zwirello2012UWB}. Later, the scheme associated with IGUULS has been thoroughly researched in recent years~\cite{Tiemann2016ATLAS, Tiemann2017Scalable, Tiemann2019ATLAS, Vecchia2019TALLA, Friedrich2021Accurate}. Like GUULS, the TDoA technique for the ranging phase and the Multilateration method for the location estimation phase is common practice in IGUULS~\cite{Zwirello2012UWB, Tiemann2017Scalable}. From a broad view, the main difference between the GUULS and IGUULS schemes is the direction of the one-way data communication flow between anchors and tags~(Fig.~\ref{fig_compare_three_uwb_schemes}). However, the said contrast ignites totally distinct usable applications~(i.e., tracking vs navigation) and several contrasting design parameters~(Section~\ref{subsec_feat_comp}).

Unlike the two unidirectional schemes~(i.e., GUULS and IGUULS), a bidirectional strategy for UWB-based localization can also be deployed~\cite{Sang2019Bidirectional}. The advantage of this approach is that the unique characteristics of the mentioned unidirectional topologies can be merged together into a single localization scheme. This enables a promising paradigm shift in the way location services are traditionally observed from the topological perspective. This is because the bidirectional localization scheme flexibly allows on-demand navigation and tracking capabilities for an application without the need to adjust the system integration process. Otherwise stated, application-specific demands for navigation or tracking purposes, as well as both conditions, can be accomplished in a single integration process using a bidirectional localization scheme. Thus, the ability of BULS to flexibly and/or elastically behave as a navigator and/or tracker on-demand offers several unique prospective applications. The term bidirectional in BULS refers to the two-way UWB signal between the anchors and the tags~(Fig.~\ref{fig_compare_three_uwb_schemes}). To the best of the authors' knowledge, Qorvo (formerly Decawave) initially presented the demonstration of bidirectional UWB localization identical to BULS in their commercial UWB development kits dubbed TREK100 Kit~(a discontinued product from the manufacturer) and MDEK1001 Kit~\cite{decawave2017MDEK1001}. Regarding the MDEK1001 kit, academic research was also carried out to improve the accuracy of location estimation of the system in~\cite{Jimenez2021Improving}. However, BULS was seldom considered as a strategic localization scheme as already stated in Section~\ref{sec_introduction}, especially in the form of academic research~\cite{Sang2019Bidirectional}. Thus, this article highlights the potential of BULS, addresses its implementation process in detail using modular processes~(Section~\ref{sec_system_model}), and discusses the prospective applications of the bidirectional scheme in other fields of research beyond UWB technology~(Section~\ref{sec_potential_beyond_uwb}).

\subsection{Feature Comparison of UWB Localization Schemes} \label{subsec_feat_comp}

Overall, Table~\ref{tab_three_uwb_systems} provides the summarized features' comparison of three UWB localization schemes discussed in Section~\ref{subsec_three_topos}. The bold letter in the Table indicates a more favorable performance measure in terms of the particular feature/properties in the table. To clarify the design parameters of the three time-based UWB localization schemes, the well-known GNSS system setup was utilized as a contrasting reference as already mentioned in Section~\ref{sec_introduction}. Correspondingly, the unidirectional GUULS system set-up is analogously defined as the GNSS-like system in this article. By contrast, the IGUULS system setup is analogous to the inverted GNSS-like system. Finally, the hybrid approach that combined both GNSS-like and its inverted topological scheme was defined as BULS, i.e., the bidirectional localization scheme in this article~(Table~\ref{tab_three_uwb_systems}).


\begin{table}
    \centering
	\caption{Feature comparison of three time-based localization schemes}
	\vspace{-0.1cm}
	\begin{tabular}{|c|c|c|c|}
			\hline
				\textbf{Properties}&\multicolumn{3}{c|}{\textbf{Time-based UWB Localization Schemes}}\\
                    \hline
				Topology & \textbf{\textit{GUULS}}& \textbf{\textit{IGUULS}} & \textbf{\textit{BULS}} \\
				& \cite{Ledergerber2015Robot, Zandian2017Robot, Santoro2023UWB, Paetru22023FlexTDOA, Yang2022VULoc, Elsanhoury2022Precision, Hamer2018Self, Corbalan2019Chorus, Grosswindhager2019SnapLoc, Santoro2021Scale} & \cite{Zwirello2012UWB, Tiemann2016ATLAS, Tiemann2017Scalable, Tiemann2019ATLAS, Vecchia2019TALLA, Friedrich2021Accurate} & \cite{Sang2019Bidirectional}\\					
				\hline
				System setup &~GNSS-like & inverted & \textbf{mixture }\\
    		       (GNSS analogy)   &    & GNSS-like & \textbf{of both}\\
				\hline		
				Ranging method &~TDoA  &~TDoA &~ToA \\		
				\hline	
				Clock  & mandatory & mandatory & \textbf{unnecessary}  \\	
                synchronization  & at anchors & at anchors &   \\
				\hline 
				Localization & navigation & tracking  & \textbf{navigation \&}  \\
				approach & purpose only & purpose only & \textbf{tracking} \\			
				\hline 
				Availability of & only at the  & only at the  & \textbf{both at the} \\
				location data & tag devices & anchor devices & \textbf{ remote \&} \\
                 & (remote side) & (central side) & \textbf{central~sides} \\
				\hline 
				Signal direction			& unidirectional & unidirectional & \textbf{bidirectional} \\
				\hline
				System-wide  & \textbf{very low} & low & medium \\		
				energy usage  &  &  &  \\		
				\hline
				No. of tags & \textbf{unlimited} & limited  & highly limited  \\
				\hline 
				No. of anchors & \textbf{limited} & \textbf{limited} & highly limited \\
				\hline 
                Multiplexing & \textbf{unnecessary} & mandatory & mandatory \\
                \hline
				UWB signal   & \textbf{active} & \textbf{active} & active both \\
                  transmission  & \textbf{anchors} & \textbf{tags} & anchors~\&~tag \\
				\hline	
		\end{tabular}          
		\label{tab_three_uwb_systems}
	\vspace{-0.5cm}
\end{table}

In terms of the ranging method~(i.e., the distance measurement phase in a time-based localization system), the two unidirectional schemes~(GUULS and IGUULS) conventionally use the TDoA technique~\cite{Ledergerber2015Robot, Friedrich2021Accurate}. In principle, the TDoA-based unidirectional schemes~(GUULS and IGUULS) require clock synchronization between anchor devices due to unavoidable clock drift errors~\cite{Zandian2017Robot, Tiemann2016ATLAS}. On the contrary, clock synchronization is not required in the system integration process of the ToA-based hybrid approach~(i.e., BULS in this article). This is due to the fact that clock drift errors can be effectively eliminated with the use of Two-Way Ranging~(TWR) method~\cite{Sang2018Analytical, Sang2019Numerical}. The ability to waive clock synchronization in BULS enables the scheme a prospective natural fit for the location estimation process beyond UWB technology, for instance, positioning in WiFi fine-time measurement~(Section~\ref{sec_potential_beyond_uwb}).  

From the topological perspective, the location information can only be retrieved on one side in GUULS and IGUULS~(Table~\ref{tab_three_uwb_systems}). This is due to the nature of unidirectional communications~(i.e., a one-way UWB signal transmission flow from a transmitter to a receiver). As a result, location data are accessible either at the tag sides in GUULS or at the anchor sides in IGUULS. Therefore, GUULS can be utilized as a navigator or for navigation purposes in location service-oriented applications~\cite{Ledergerber2015Robot, Zandian2017Robot}. On the other hand, IGUULS can be utilized as a tracking scenario or a surveillance/monitoring system in location service-oriented applications~\cite{Tiemann2017Scalable, Vecchia2019TALLA}. In contrast, BULS exhibits the capability that can flexibly and/or elastically combine the mentioned two unique perspectives of localization~(i.e., navigation and tracking scenarios) using just a single scheme~\cite{Sang2019Bidirectional}. Consequently, in BULS, location information is available at both the tag and anchor sides. This flexible capability opens up several potential use cases in location-based services~(Section~\ref{subsec_potential_usecases_buls}).

In connection with the system-wide energy consumption, the GUULS~(the GNSS-like topology) is considerably more efficient than the IGUULS~(the inverted GNSS-like topology) and BULS~(Table~\ref{tab_three_uwb_systems}). This is due to the fact that only the anchors are accountable for transmitting the UWB signals. In practice, the power dissipation~(consumption) of a UWB hardware chip in the signal transmission process is considerably larger than the receiving as well as the idle processes~\cite{Zandian2019Ultra}. In theory, the number of anchors to be used for the UWB localization system is significantly less than the number of tags in the system. In contrast, the demand for system-wide power consumption of IGUULS is getting high as the number of tags in the system increases. This is due to the fact that tags are responsible for UWB signal transmission in IGUULS~\cite{Tiemann2017Scalable, Vecchia2019TALLA}. By comparison with GUULS and IGUULS, the system-wide energy consumption of BULS is the highest among the three time-based UWB topological schemes because both anchors and tags are involved in the UWB signal transmission for the location estimation process~(Table~\ref{tab_three_uwb_systems}). 

Overall, the most notable strength of GUULS is that the number of tags in the system can be scaled up to the infinity~\cite{Santoro2021Scale, Santoro2023UWB, Paetru22023FlexTDOA, Yang2022VULoc}~(i.e., there is no limit for the number of tags in it). Moreover, the multiplexing scheme is generally not required in the system integration process of GUULS~(Table~\ref{tab_three_uwb_systems}). This empowers the GUULS topology as a simple architecture with a highly scalable localization system~(i.e., unlimited tags). However, the weakness of GUULS is that it can be used only for navigation purposes~(for instance, self-localization of a robot). This limits the applicable areas of GUULS, especially in indoor environments. In contrast, the multiplexing scheme is mandatory for both IGUULS and BULS in order to prevent possible signal collisions and guarantee the prescribed access time slot for each tag in the system~\cite{Vecchia2019TALLA, Sang2019Bidirectional}. The exceptional strength of IGUULS is that the scheme can be used in the tracking scenario, which has several practical applications including logistics, industrial automation, safety measures in GNSS-deprived zones, etc. For that reason, the research on the system integration process of IGUULS  in applications has risen in recent years~\cite{Zwirello2012UWB, Tiemann2016ATLAS, Tiemann2017Scalable, Tiemann2019ATLAS, Vecchia2019TALLA, Friedrich2021Accurate}. 

The main hurdle in the topology of GUULS and IGUULS is the requirement of a system-wide clock synchronization~(Table~\ref{tab_three_uwb_systems}). 
On the contrary, the BULS relaxes the intricate and complex synchronization process by using the TWR method. Moreover, the use of TWR enables the BULS to behave as a navigator and a tracker on-demand in applications. Thus, BULS combines the core advantageous features of GUULS~(navigation process) and IGUULS~(tracking process) within one system set-up. The flexible and elastic nature of the BULS opens endless potential applications as well as a paradigm shift in the way location services are observed. Moreover, the system integration process of BULS can easily be transferred into other similar technologies, e.g., Wi-Fi fine time measurement system defined in IEEE 802.11~mc~\cite{Guo2019Indoor, Ibrahim2018Verification, Ma2022WiFi}. This is because TWR is naturally fitted to the data communication process involved in the said Wi-Fi system, e.g. the use of Carrier Sense Multiple Access~(CSMA) protocol. However, the BULS also comes with a cost in terms of scalability due to the need for more ranges in the TWR~(Section~\ref{subsec_ranging_process}) as well as the multiplexing process~(Section~\ref{subsec_multiplexing_process}). Theoretically, the BULS shows a considerably limited impact in terms of scalability compared to its counterpart the unidirectional localization schemes~(i.e., GUULS and IGUULS).

\subsection{Potential Use-cases of Bidirectional Localization Scheme} \label{subsec_potential_usecases_buls}

In general, GUULS and IGUULS are useful for large-scale navigation and tracking application respectively. Besides, the GUULS and IGUULS are specifically designed for high-time resolution technology like UWB. Thus, the schemes cannot be easily migrated into application areas in other technologies such as WiFi, and the 5G/6G mobile networks. One of the major stumbling blocks is the strict requirement of clock synchronization in the system integration process for GUULS and IGUULS, as mentioned in the previous section. 

By contrast, the BULS can be re-implemented with other time-based wireless radio technologies without the need to change the core integration procedures. For instance, one of the prominent candidates regarding such technology includes the WiFi fine-time technique specified in the recent amendment of IEEE~80.11mc~\cite{Ibrahim2018Verification, Guo2019Indoor, Ma2022WiFi}~( Section~\ref{sec_potential_beyond_uwb}). In principle, the functional use cases of the bidirectional localization scheme addressed in this section are not limited to the UWB technology alone, though it is the primary focus of this article. Instead, the scheme can be utilized in many other fields, as discussed a couple of them in the coming Section~\ref{sec_potential_beyond_uwb}. 

For motivational purposes, the prospective use cases of the BULS are highlighted in this section. To mention a few, the BULS can be used in navigation and tracking for the underground car parking system, player tracking and performance analysis in sports, precise location-oriented factory automation for small to medium sizes, etc. Moreover, BULS can also be used in mission-critical safety measures, such as precise navigation and tracking scenarios for firefighters, underground mining, and underwater exploration. Besides, BULS possesses the potential features for the location estimation process in 5G and beyond technologies. This is because BULS naturally fits into the communication protocols utilized in 5G/6G networks, especially for guaranteed reliability. In addition, the customizable personalized and dedicated networks established in 5G/6G are promising features for location estimation services primarily for small and medium-scale factories and other environments. BULS is a natural fit for the precise location estimation process in those mentioned areas with the minimum system migration process.

\section{Viable Prospects of the Bidirectional Localization Scheme Beyond UWB} \label{sec_potential_beyond_uwb}

Though the primary focus of this article is on UWB technology, the bidirectional localization scheme is not bounded to the UWB-based positioning systems alone. In fact, the scheme can be applied to other location-oriented services in different fields. For instance,~WiFi Fine Time Measurement~(FTM) was recently adopted in the IEEE 802.11~mc standard~\cite{Ibrahim2018Verification, Guo2019Indoor, Han2019Smartphone}. For location estimation, the distance measuring process of WiFi FTM relies on TWR method~\cite{Ibrahim2018Verification, Guo2019Indoor, Nkrow2022WiFi, Han2019Smartphone}. In general, TWR is considered the core enabler of the bidirectional scheme as it inherently enables communication flow between two wireless transceivers by exchanging the perceived data mutually~(Section~\ref{subsec_minimum_setup}). Thus, the bidirectional localization scheme can be considered as a prospective positioning topology for Wi-Fi FTM in the location estimation processes as well as the applications enabled by the technology.  

\begin{figure}[!t]
\centering
\includegraphics[width=8.8cm]{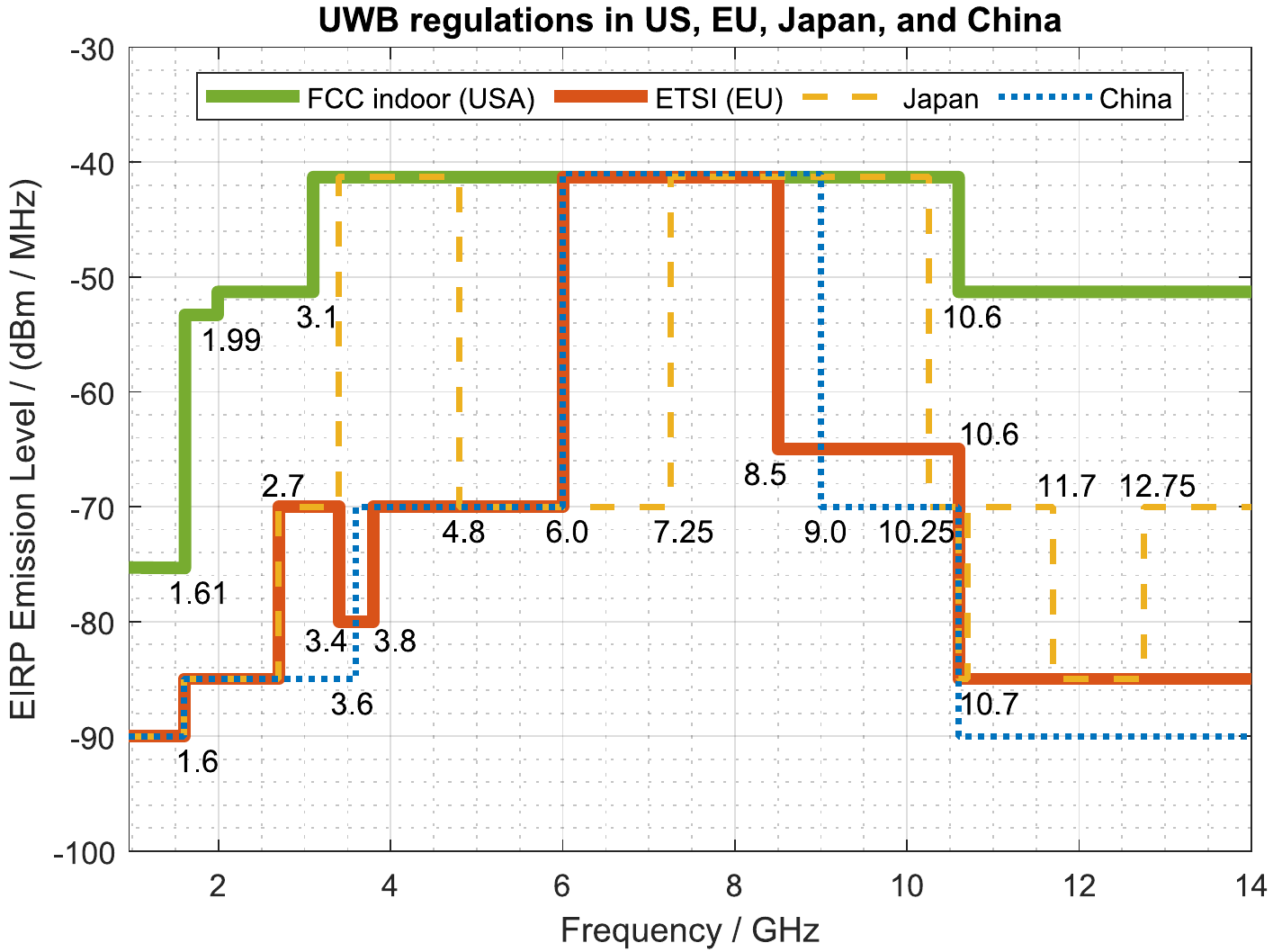} 
\vspace{-0.5cm}
\caption{Regulation of UWB spectrum masks applied in the US, EU, Japan, and China \cite{Sang2022Dissertation}. The details of the presented spectrum masks and regulations for other countries can be observed in~\cite{ETSI2019Short}.} 
\label{fig_uwb_regulation}
\vspace{-0.4cm}
\end{figure}

Likewise, the bidirectional localization scheme offers several potential location-aware applications in 5G and beyond wireless communications. For instance, the viable scenarios of the bidirectional localization scheme in 5G/6G include:~(i) real-time in-networking positioning system for swarm robotics, (ii) safety aware location estimation in vehicle platooning system, (iii) real-time positioning system in the internet of drones, (iv) localization for indoor car parking scenario, (v) self-navigation process for the driverless transport systems in GNSS-deprived zones, (vi) precise real-time positioning system for cooperative autonomous vehicles, etc. 

In principle, frequency spectrum allocations for 5G and beyond are still under active discussions and research, especially for the use-case of 5G positioning system~\cite{Dwivedi2021Positioning}. From the perspective of UWB spectrum allocation, Fig.~\ref{fig_uwb_regulation} illustrates the current restriction of UWB usage in terms of the Effective Isotropic Radiated Power~(EIRP) mask regulated by four regulatory bodies, namely FCC in the US, the European Telecommunications Standards Institute (ETSI) in EU, Japan, and China. The figure specifically indicates the spectrum usage of UWB technology, which also collides with the spectrum of interest in 56/6G wireless networks~\cite{Yang2021Spectrum}. It is interesting to see that the frequency spectrums of the common maximum allowable EIRP for the depicted four regulators are between \SI{7.25}{\GHz} and \SI{8.5}{\GHz}~(Fig.~\ref{fig_uwb_regulation}). Except for Japan, the  maximum allowable EIRP is~\SI{6.0}{\GHz} to \SI{8.5}{\GHz}. Indeed, the mentioned spectrum interval of the maximum allowable EIRP is literally the same for all regulators around the world at the time of writing this article~(see the detailed spectrum restriction of UWB across the globe in~\cite{ETSI2019Short}). In particular, the exact value of the maximum allowable EIRP is specified as \SI{-41.3}{\dBm} for UWB worldwide except in China, where it is specified as \SI{-41.0}{\dBm}~\cite{ETSI2019Short}~(Fig.~\ref{fig_uwb_regulation}). 

Regarding this, the 5G and beyond wireless networks are designed for service-oriented customizable tactical and technological entities and applications~\cite{Ghosh20195G}. Therefore, new methodologies of spectrum-sharing schemes are paramount to the efficient usage of limited resources in 5G/6G~\cite{Yang2021Spectrum}. As a consequence, the adaptation and/or allocation of the currently accessible UWB spectrum into the 5G/6G network is one of the beneficial approaches to meet the necessities of some targeted applications, especially for location-oriented solutions in the areas of industrial and vehicular automation systems. Taking into account the perspectives expressed in this and previous sections, the system model of the bidirectional localization scheme based primarily on UWB technology (i.e. bidirectional UWB localization) is presented in Section~\ref{sec_system_model}. 

\vspace{1cm} 
\section{System Model and the Integration Processes of Bidirectional UWB Localization} \label{sec_system_model}

\begin{figure}[!t]
    \centering
    \includegraphics[width=8.8cm]{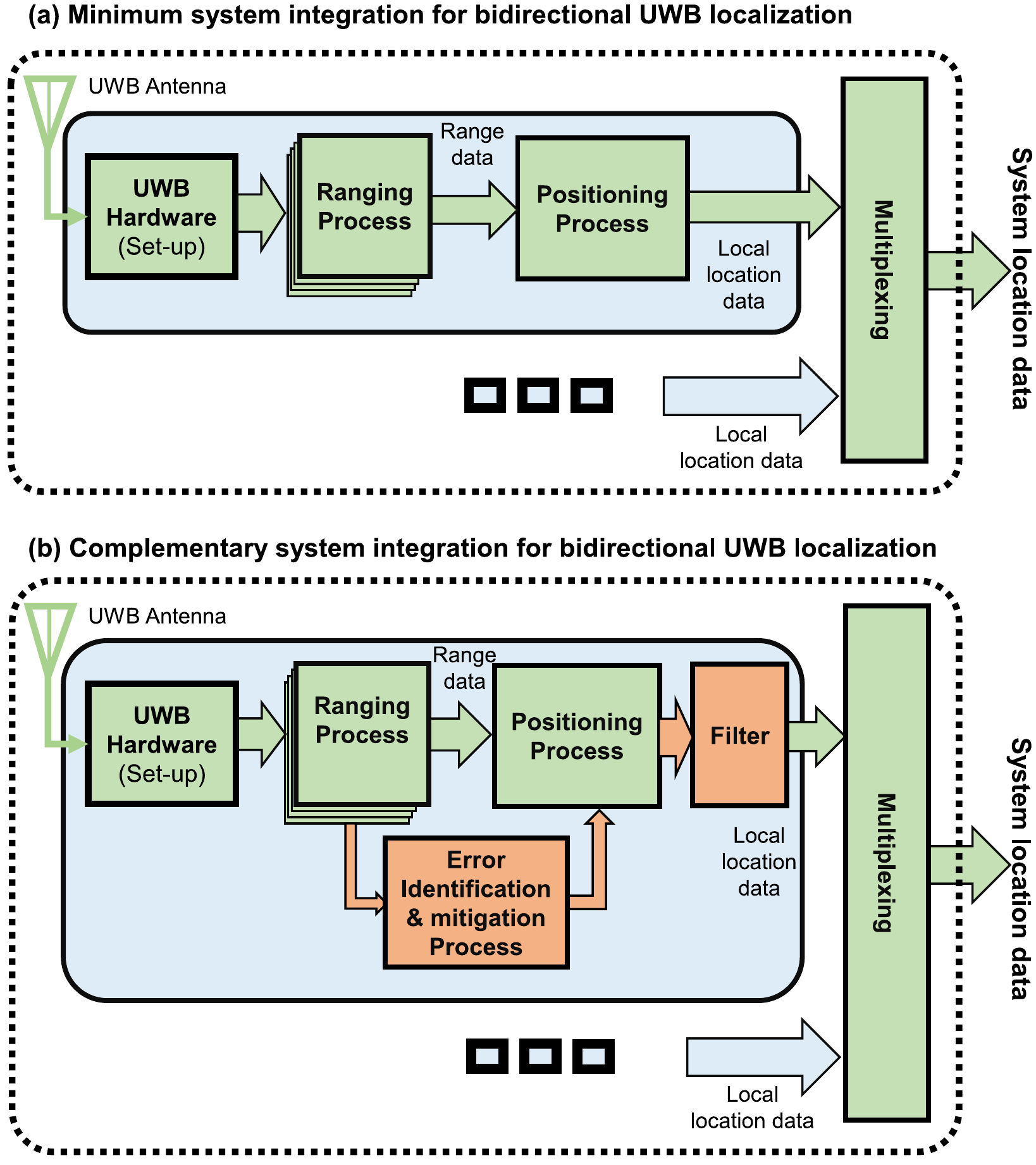}
    \vspace{-0.5cm}
    \caption{Block diagram of two system integration processes for bidirectional UWB localization:~(a) a minimum set-up for the system integration process, (b)~a complementary set-up for the system integration process.}
    \label{fig_system_setup_block_diag}
    \vspace{-0.3cm}
\end{figure}

This section elaborates on the system model of BULS, which is also viewed as an elastic positioning scheme for GNSS-deprived zones in this article. The modular design principle was taken into account in this section to present the integration processes of each operational block associated with the BULS~(Fig.~\ref{fig_system_setup_block_diag}). In other words, the system integration process of BULS was seen as the combination of several modules, which are independent and can be composed of different methodologies based on the requirements. Each of the operational modules is depicted as separate blocks in~Fig.~\ref{fig_system_setup_block_diag}. In this regard, two system integration processes were specifically addressed in this article and termed the minimum system setup for BULS~(Section~\ref{subsec_minimum_setup}) and the complimentary system setup~(Section~\ref{subsec_complementary_setup}). In short, BULS can be considered a complete localization system with minimum requirements if the four operational modules depicted in Fig.~\ref{fig_system_setup_block_diag}~(a) are established. In contrast, the modular blocks colored with light orange in Fig.~\ref{fig_system_setup_block_diag}~(b) are beneficial for the enhancement of the system. Hence, these modules are recognized as a complementary process and addressed in Section~\ref{subsec_complementary_setup}.  

The modular approach simplifies the description of the integration procedure of BULS. However, it should be noted that each functional module preserves its own standalone and extensible characters composed of different methodologies. For instance, there are several algorithms~\cite{Sang2019Comparative} that can be used as the location estimation method in BULS regarding the functional module annotated as position process~(Fig.~\ref{fig_system_setup_block_diag}). The same applies to other functional modules such as the ranging process, multiplexing, identification, etc. In this article, the detailed applicable methods in each module are ignored for simplicity. Instead, one methodology for each module is chosen to describe the integration process. Indeed, the chosen one in the article can be replaced with other comparable methods by preferences. In light of this, a brief on the practical implementation process is also discussed in~Section~\ref{subsec_digest_practice_impl}.

\subsection{Model Integration of BULS with Minimum System Requirements} \label{subsec_minimum_setup}

In general, there are four operational modules mandatory for the model integration of BULS with minimum setup~(Fig.~\ref{fig_system_setup_block_diag}~(a)). Those four operational modules are (i) the UWB hardware components, (ii) the ranging process, (iii) the location estimation process, and (iv) the multiplexing process of the system as concisely discussed in subsequent subsections.

\subsubsection{Typical Hardware Components for UWB Localization Systems} \label{subsec_hardware_components}

In general, the typical hardware of a UWB device is composed of at least four main components. The first component corresponds to the antenna of the UWB device, which is utilized for the transmission and receiving process of UWB radio signals. The second principal component in the hardware is the UWB chip, generally produced by a manufacturer. The chip specification is usually defined in compliance with a standard such as IEEE~802.15.4z~\cite{std2020IEEE} (i.e., the latest amendment to the 802.15.4-2020 standard on the UWB specification at the time of writing this article). The major tasks of the UWB chip include precise time-stamping,~UWB signal processing, the configuration of the~UWB channel, data rate, emission power, as well as other suitable parameters based on the intended application. The third principal component of the hardware is the central processing unit which is responsible for controlling the data communication flow in the UWB chip. Typically, it is a microcontroller unit in many applications. The fourth core component is a  high-precision oscillator, whose primary task is to provide an accurate tick to the local clock of the UWB hardware. The high-precision clock in UWB hardware is paramount for achieving accurate timestamps used for the ranging process in UWB localization, especially for the TDoA-based unidirectional approaches, i.e., referring to the GUULS and IGUULS discussed in Section~\ref{subsec_feat_comp}.

On the whole, the underlying hardware components of a UWB device are principally the same for the three defined time-based topological schemes~(GUULS, IGUULS, and BULS). However, it needs to be cautious that the high precision oscillators are mandatory for the TDoA-based unidirectional schemes~(GUULS and IGUULS) due to the stringent necessity of clock synchronization in the system integration~(Section~\ref{subsec_feat_comp}). In contrast, BULS, which is based on the TWR method, can operate efficiently just by using an ordinary low-cost oscillator~\cite{Sang2019Numerical}. Today, UWB chips are available at a very low cost in the electronic markets thanks to the manufacturers like Decawave~(at present under Qorvo), Ubisense, Bespoon, NXP, etc. Moreover, smartphone manufacturers such as Apple and Samsung have already integrated~UWB chips into their products~\cite{Coppens2022Overview}. It is expected that the exponential growth of~UWB chips in the market and the rise of the technology will be evident in the near future.

\subsubsection{Ranging Process for Bidirectional UWB Localization} \label{subsec_ranging_process}

The methodology used in the ranging process basically determines the applicable localization scheme~(i.e., either GUULS, IGUULS, or BULS) in the system or vice versa. This is because the applicable ranging techniques are closely tied up with the type of signal measurement processes, i.e., TDoA, ToA, AoA, or RSS~\cite{Yassin2017Recent} as expressed in Section~\ref{subsec_time_based_loc}. The UWB localization systems based on the unidirectional approaches~(GUULS and IGUULS) are commonly built upon the TDoA techniques as mentioned before. 
On the contrary, the UWB localization system based on the bidirectional approach is fundamentally established on the ToA technique~\cite{Sang2019Bidirectional}. 

\begin{figure}
    \centering
    \includegraphics[width=8.8cm]{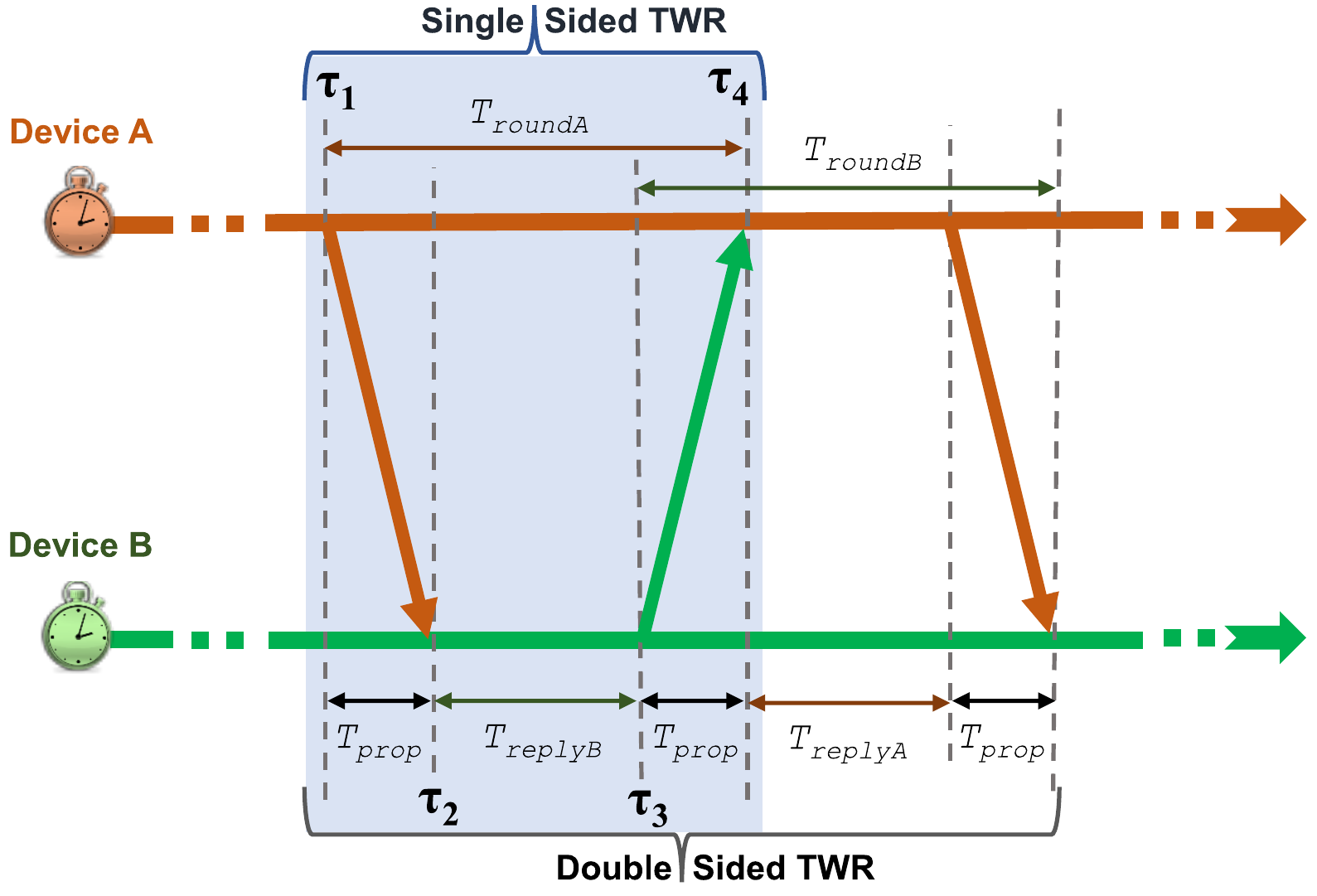}
    \vspace{-0.5cm}
    \caption{Expression of TWR method in the ranging process of the BULS~\cite{Sang2022Dissertation}. The image shows both SS-TWR and DS-TWR on an image.} 
    \label{fig_twr_scheme}
    \vspace{-0.5cm}
\end{figure}

It is noteworthy that the main enabler of BULS stems from the use of a TWR technique~(Fig.~\ref{fig_twr_scheme}) in the distance measurement process of the system. The measured distances using TWR are commonly termed as true range measurements~\cite{Sang2019Bidirectional} as opposed to pseudo-ranges in GNSS and alike. This is due to the fact that TWR is capable of eliminating or reducing clock drift errors~\cite{Sang2018Analytical}. Besides, a recent study in~\cite{Chen2022PnPLoc} demonstrated the crucial role of TWR in the system integration process of a plug-and-play UWB-based localization system. 
Regarding this, the authors' previous works~\cite{Sang2018Analytical, Sang2019Numerical} revealed that an Alternative Double-sided TWR~(AltDS-TWR) is the most consistent and reliable ranging method among other available TWRs in the literature. AltDS-TWR was originally proposed in~\cite{Neirynck2016Alternative} and we keep the initially given name by the creators of the method in this article. 
Accordingly, AltDS-TWR has been later adopted as the default Double-sided TWR~(DS-TWR) technique for range-based measurements in IEEE 802.15.4z-2020 standard~\cite{std2020IEEE}. For more details on AltDS-TWR as a preferred method compared to other TWR techniques in the literature, we refer to our previous work~\cite{Sang2019Numerical, Sang2018Analytical}. 

For illustrative purposes, Fig.~\ref{fig_twr_scheme} depicts two basic TWR methods namely SS-TWR and DS-TWR (i.e., AltDS-TWR in this article) respectively. The SS-TWR~(shaded area in Fig.~\ref{fig_twr_scheme}) can be formulated as~\cite{Sang2018Analytical}:
\begin{equation}
\label{eq_ss_twr}
T_{prop} = \dfrac{1}{2} \cdot (T_{roundA} - T_{replyB})
\end{equation}
where $T_{prop}$ is the propagation time or TOF of the UWB signal between two transceivers, $T_{roundA} = \tau_{4} - \tau_{1} $ is the round-trip time of a signal measured at the local clock of Device A, and $T_{replyB} = \tau_{3} - \tau_{2}$ is the reply time of a signal measured at the local clock of Device B.

In the same manner, the formulation of DS-TWR~(Fig.~\ref{fig_twr_scheme}) specifically AltDS-TWR~(i.e., the name given by the original creator of the method~\cite{Neirynck2016Alternative}) can be formulated as~\cite{Sang2019Numerical, std2020IEEE}:
\begin{equation}
\label{eq_ds_twr_altds}
T_{prop} = \dfrac{T_{roundA} \cdot T_{roundB} - T_{replyA} \cdot T_{replyB}} {T_{roundA} + T_{roundB} + T_{replyA} + T_{replyB}} 
\end{equation}

where $T_{roundA}$ and $T_{roundB}$ are the round-trip times of a signal measured at Device A and B, respectively. $T_{replyA}$ and $T_{replyB}$ are the reply times or response times measured at Device A and B, respectively.

\subsubsection{Location Estimation Process} \label{subsec_positioning_process}

\begin{figure}
    \centering
    \includegraphics[width=8.8cm]{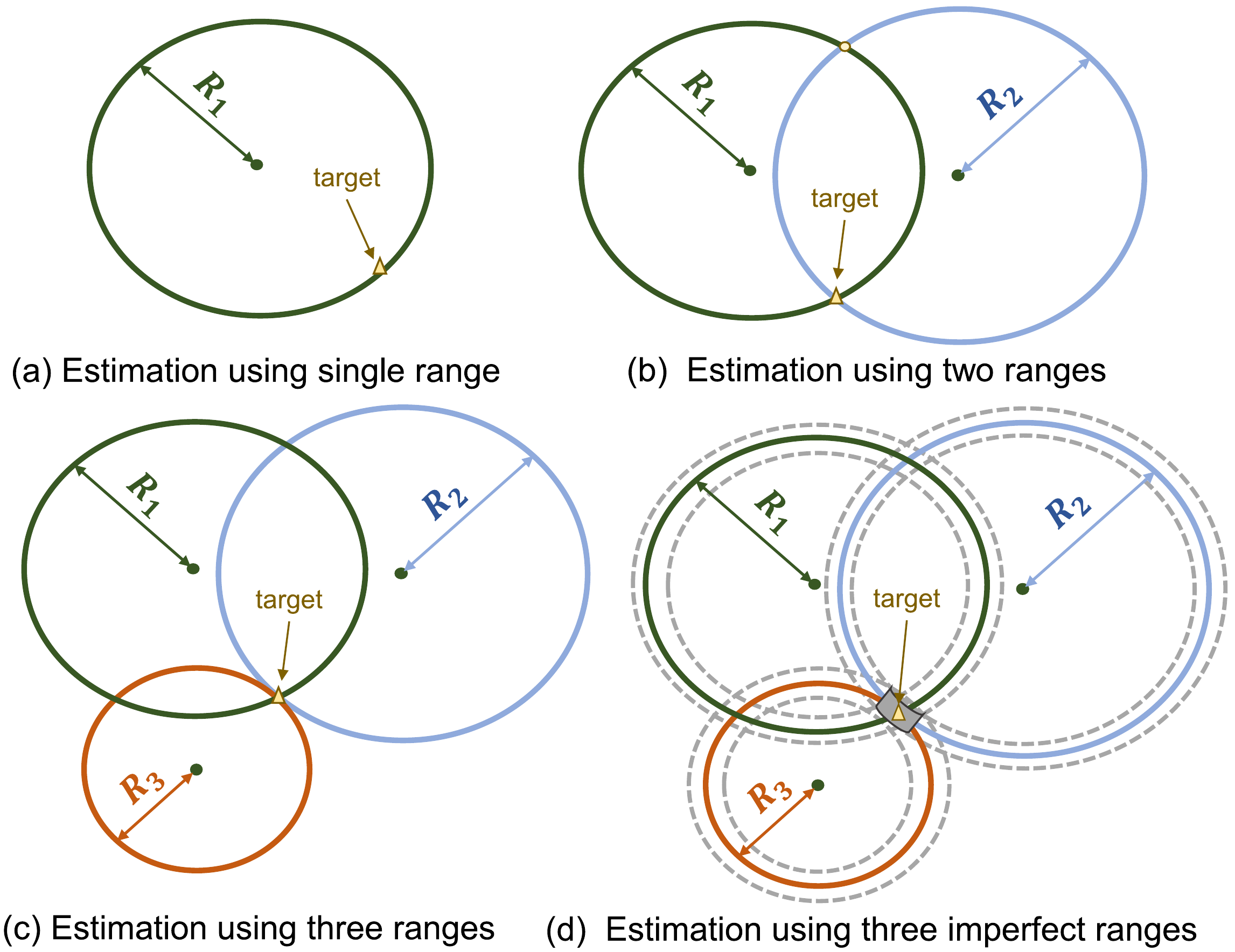}
    \vspace{-0.5cm}
    \caption{Demonstration of the location estimation process in 2D using true-range measurements~\cite{Sang2022Dissertation}. The estimation of the target location is based on:~(a) a single range, (b)~two ranges, (c)~three ranges, and (d)~three imperfect ranges.}
    \label{fig_2D_positioning}
    \vspace{-0.3cm}
\end{figure}

\begin{figure}
    \centering
    \includegraphics[width=8.8cm]{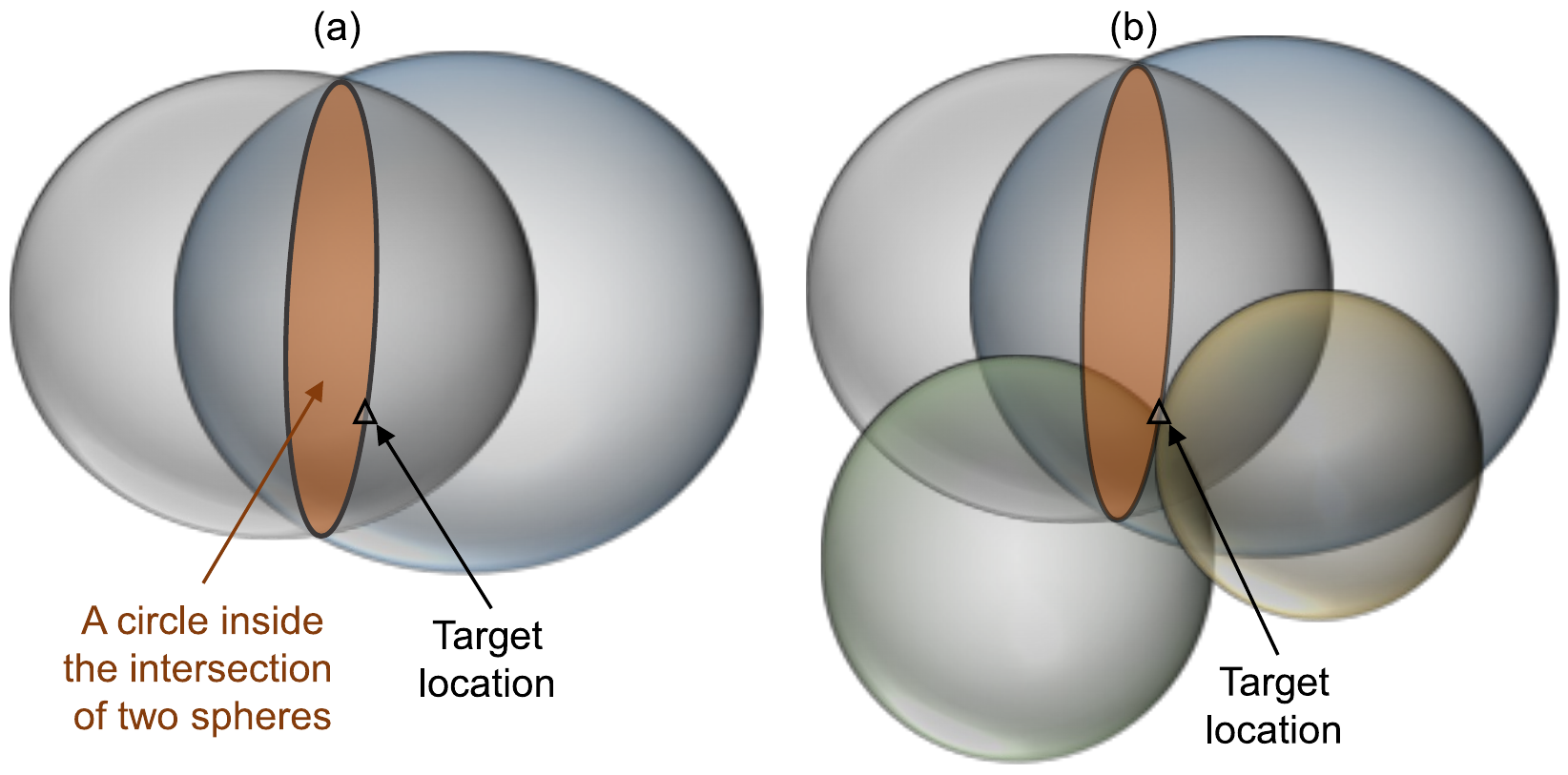}
    \vspace{-0.5cm}
    \caption{Illustration of the location estimation process in 3D assuming an ideal case with perfect ranges:~(a) estimation of the target location in 3D using two spheres, (b) estimation of the exact target location in 3D using four spheres.}
    \label{fig_3D_positioning}
    \vspace{-0.5cm}
\end{figure}

The determination of a target location in BULS is basically based on the measured distances achieved in the ranging process and a positioning algorithm. In the UWB localization system, a positioning algorithm requires at least three ranges to estimate a target location in 2D space~(Fig.~\ref{fig_2D_positioning}) and at least four ranges in 3D space~(Fig.~\ref{fig_3D_positioning}). Normally, eight or more anchors are utilized for precise location estimation processes in practice, for instance, a player tracking system in Sports using UWB localization~\cite{BastidaCastillo2019Accuracy}. For brevity and clarity, this article addresses the fundamental of the lateration-based positioning method using true-ranges~(Fig.~\ref{fig_2D_positioning} and Fig.~\ref{fig_3D_positioning}). In short, lateration is termed Trilateration if the localization system uses exactly three ranges in the location estimation process. In contrast, it is called Multilateration if four or more ranges are involved in the estimation process. 


The necessity of at least three ranges in the 2D setup and at least four in the 3D can be visualized with a circular and spherical representation as depicted in~Fig.~\ref{fig_2D_positioning} and Fig.~\ref{fig_3D_positioning}, especially for the system that used true-ranges in the signal measurement process. For 2D, the estimation using a single range alone for a target is anywhere on a circle~(Fig.~\ref{fig_2D_positioning}~(a)). The feasibility of the target location is reduced to two points at the intersection of two circles if two different ranges from two references are utilized in the estimation process~(Fig.~\ref{fig_2D_positioning}~(b)). Eventually, the exact position of a target can be determined in 2D space at the intersection point of three circles by using three ranges~(Fig.~\ref{fig_2D_positioning}~(c)). Analogously, the estimation using a single range in 3D space is a sphere. The use of two ranges minimizes the estimation of a location into a circle because the intersection between two spheres is a circle~(Fig.~\ref{fig_3D_positioning}~(a)). Therefore, at least four ranges are mandatory for location estimation in 3D space~(Fig.~\ref{fig_3D_positioning}~(b)). 

However, the measured ranges are not ideal in practice and normally affected by noises such as None-Line-of-Sight~(NLOS) conditions and multi-path scenarios in the propagation of radio signals~\cite{Dardari2009Ranging}. In consequence, the estimation of the target cannot be, in most cases, determined at an exact single point in practice~(e.g., see the 2D case in~Fig.~\ref{fig_2D_positioning}~(d)). Therefore, there are several location estimation algorithms for UWB-based localization systems based on closed-form methods, iterative approaches as well as statistical techniques~\cite{Sang2019Comparative, Shen2008Performance}. Specifically, our previous work in~\cite{Sang2019Comparative} and the dissertation in~\cite{Sang2022Dissertation} addressed five positioning algorithms commonly used in UWB localization using real-world experimental data in the evaluations. Additionally, the associated research data as well as the source codes for the mentioned research studies were openly provided in~\cite{Sang2022Repo}. For demonstration purposes, the positioning algorithm based on the Taylor series technique is highlighted and explained in this article. 

In general, the equation related to true-range measurement in wireless communication can be defined in 3D space as follows~\cite{Frattasi2017Mobile}~(Fig.~\ref{fig_2D_positioning} and Fig.~\ref{fig_3D_positioning}):
\begin{equation}
d_{i}^2 = (x_{i} - x_{t})^2 + (y_{i} - y_{t})^2 + (z_{i} - z_{t})^2 \label{eq_range_3d}
\end{equation}
where, $d_i$ is the measured distance~(range) between the anchor $X_{i}$ and tag $X_{t}$. In \eqref{eq_range_3d}, it is assumed that the tag~(target device) is located at position $X_{t} = [x_t, y_t, z_t]^T$ in 3D space whereas the fixed anchors' location~(the centers of spheres) are at $X_{i} = [x_i, y_i, z_i]^T$. $i = 1, 2, ..., N$ are the identities of the anchors.

Based on the spherical ranging equation given in~\eqref{eq_range_3d}), a function that corresponds to the $i$th measurement between the $i$th anchor and a tag can be defined as~\cite{FOY1976Position}~(Fig.~\ref{fig_2D_positioning} and Fig.~\ref{fig_3D_positioning}):
\begin{eqnarray} 
	f_i (x, y, z) &=& \sqrt{(x_i - x)^2 + (y_i - y)^2 + (z_i - z)^2} \label{eq_TS_func} \\
	&=& d_i + \varepsilon _i ~~~~~~~(i = 1, 2, ..., n) \nonumber	
\end{eqnarray}
where $\varepsilon_i$ is the range estimation error between a tag and the $i$th anchor. In this regard, it is assumed that the errors~($\varepsilon$) are statically distributed, and its elements are independent of each other based on zero-mean Gaussian random variables~\cite{FOY1976Position}. Thus, the error covariance matrix can be written as:
\begin{equation}
R = E[\varepsilon \varepsilon^T] = diag[\sigma^2  ... \sigma^2] \label{weight_TS}
\end{equation}
where $\sigma$ is the range estimation error.

If we suppose $(x_v, y_v, z_v)$ is an initial guess of the true tag's location $(x_t, y_t, z_t)$, we can express as:
\begin{equation}
x_t = x_v + \delta_x, y_t = y_v + \delta_y, z_t = z_v + \delta_z  \label{full_location}
\end{equation} 
where, $\delta_x$, $\delta_y$, and $\delta_z$ are the location errors (i.e., the incremental error between the guess and the ground truth or true position) of a tag to be determined.

If Equation~\eqref{eq_TS_func} is expanded into the Taylor-series expansion by keeping the first-order term, it can be expressed as in the following:
\begin{equation}
f_{i,v} + a_{i,1} \cdot \delta_x + a_{i,2} \cdot \delta_y + a_{i,3} \cdot \delta_z \approx d_i + \varepsilon_i    \label{1st_order_TS_form}
\end{equation}
where, 
\begin{equation}
f_{i,v} = f_i(x_v, y_v, z_v),
~~ a_{i,1} = \frac{\partial f_i}{\partial x}\Bigr|_{\substack{x_v, y_v, z_v}} = \dfrac{x_v - x_i}{r_i}, \nonumber
\end{equation}
\begin{equation}
a_{i, 2}=\dfrac{\partial f_i}{\partial y}\Bigr|_{\substack{x_v, y_v, z_v}}=\dfrac{y_v - y_i}{r_i}, a_{i,3}=\frac{\partial f_i}{\partial z}\Bigr|_{\substack{x_v, y_v, z_v}} = \dfrac{z_v - z_i}{r_i}, \nonumber
\end{equation}
\begin{equation}
r_i =\sqrt{(x_i - x_v)^2 + (y_i - y_v)^2 + (z_i - z_v)^2} \nonumber
\end{equation}
Equation~\eqref{1st_order_TS_form} can be written in matrix notation as
\begin{equation}
H \delta = \Delta d + \varepsilon  \label{matrix_form_TS}
\end{equation}
where, $\Delta d = d_i - f_{i,v} = d_i - r_i$,
\[ H = \begin{bmatrix} 
a_{1,1} & a_{1,2}  & a_{1,3} \\
a_{2,1} & a_{2,2}  & a_{2,3}  \\
&    ...         &               \\
a_{n,1} & a_{n,2}  & a_{n,3} 
\end{bmatrix},
\quad
\delta = \begin{bmatrix}
\delta_x \\
\delta_y \\
\delta_z
\end{bmatrix},
\quad
\varepsilon = \begin{bmatrix}
\varepsilon_1 \\
\varepsilon_2 \\
... \\
\varepsilon_n
\end{bmatrix}
\]
Using the covariance of measurement error~($R$) in~\eqref{weight_TS} as a weight, \eqref{matrix_form_TS} can be solved using the overdetermined weighted least squares method~\cite{FOY1976Position} as:
\begin{equation}
\delta = (H^T R^{-1} H)^{-1}  H^T R^{-1} \Delta d   \label{incremental_delta}
\end{equation}

Finally, the location of the target tag device~($x_t, y_t, z_t$) can be estimated by using an iterative approach~(i.e., continually refining the error in each process as new data are received in the measurement), if we substitute the computed incremental error~($\delta$) from~\eqref{incremental_delta} to~\eqref{full_location}. 

In summary, the location estimation in the Taylor-series technique starts with an initial guess. Then, the method improves its estimation at each iteration step by determining the solution of the local linear least-squares parameters. The method is categorized under the iterative approach because the beginning stages of the positioning algorithm will encounter some errors depending on how well the initial guess comprises the actual ground truth. As the iteration continues, the estimation of the algorithm will eventually converge to the pragmatic actual solution. However, it is important that a realistic initial guess should generally be provided to the system to avoid probable divergence in iterative approaches.

\subsubsection{Multiplexing Process} \label{subsec_multiplexing_process}

\begin{figure}[!t]
    \centering
    \includegraphics[width=8.8cm]{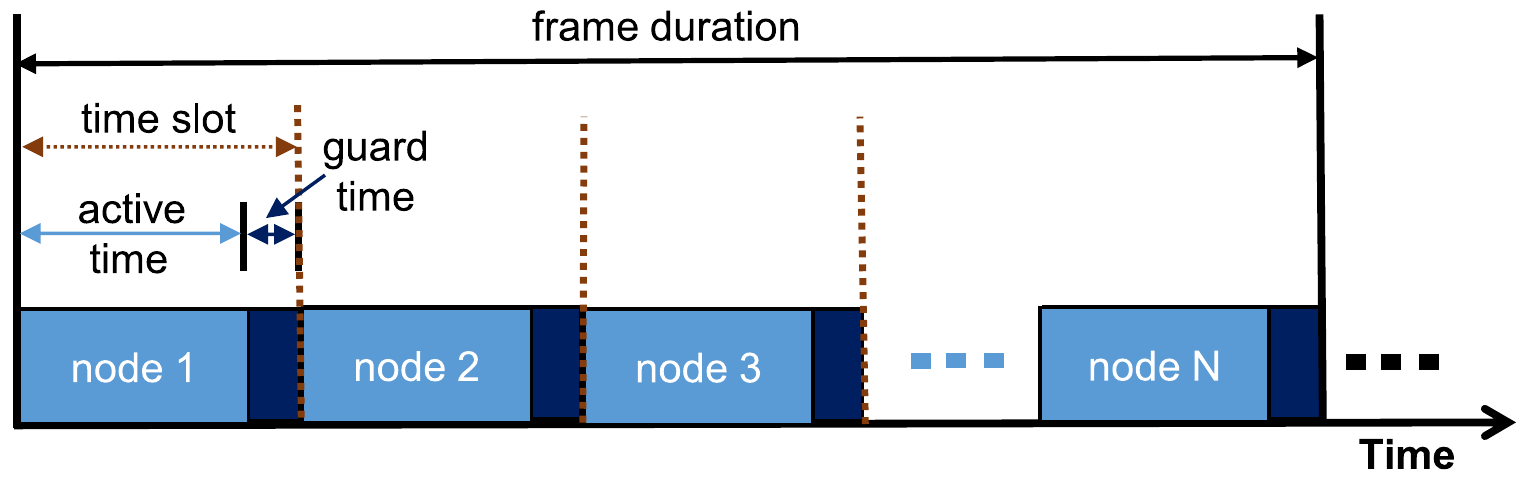}
    \vspace{-0.7cm}
    \caption{TDMA-based multiplexing scheme with a bounded system latency~(i.e., frame duration) and individual time slots for each node.}
    \label{fig_tdam_scheme}
    \vspace{-0.5cm}
\end{figure}

Multiplexing is a method that combines multiple individual outcomes into a single system output over a shared medium. The primary goal of multiplexing is to share limited resources within a system by avoiding signal collisions while achieving efficient processing time. Based on the academic papers and technical documents published in the literature, there exists only one multiplexing scheme by far commonly applied in UWB-based localization systems. That method is a Time Division Multiple Access~(TDMA) technique~\cite{Tiemann2016ATLAS, Sang2019Bidirectional, Vecchia2019TALLA}~(Fig.~\ref{fig_tdam_scheme}). Other multiplexing approaches, such as frequency, code division, and hybrid techniques, are not utilized especially for the UWB-based system integration processes. One of the primary reasons is that the TDMA scheme is very appealing for the high temporal resolution produced by IR-UWB technology~\cite{Win2009History}. 

In short, the TDMA multiplexing scheme is beneficial for a system that requires power efficiency~(e.g., no idle listening) as well as guaranteed message delivery without collisions~(i.e., each node possesses its own time slot for data processing as depicted in Fig.~\ref{fig_twr_scheme}). A single time slot in the TDMA scheme is generally composed of an active time slot~(i.e., the actual working duration of a node) and a guard time~(i.e., a reserved time slot to compensate for errors/delays due to small variations during processing) as shown in Fig.~\ref{fig_tdam_scheme}. The drawbacks of TDMA, however, are the requirement of a bounded system latency~(i.e., a constraint predetermined time which is denoted as frame duration in Fig.~\ref{fig_tdam_scheme}) and the waste of resources for unused time slots in the system. Moreover, network-wide synchronization is mandatory for TDMA in order to eliminate time-jittering effects, which are a small variation of time/delay naturally encountered in time-restricted system integration process~\cite{Sang2019Bidirectional, Ganeriwal2003Timing}. 

For a system that uses TWR schemes~(e.g., BULS), the node synchronization required for TDMA can be implemented easily. One such algorithm is a~\textit{timing-sync protocol} originally proposed for wireless sensor networks in~\cite{Ganeriwal2003Timing}. The fundamental formula for the protocol is as follows~(i.e., referred to the shaded area in~Fig.~\ref{fig_twr_scheme} for the illustration of the concept):
\begin{equation}
\Delta e = \dfrac{(\tau_{2} - \tau_{1}) - (\tau_{4} - \tau_{3})}{2} \label{eq_tdma_clockDrift}
\end{equation}
where $\Delta e$ symbolizes the clock drift between two wireless devices, $\tau_{1}$ and $\tau_{4}$ are the transmitted and received timestamps of device A respectively, and $\tau_{3}$ and $\tau_{2}$ are the transmitted and received timestamps of device B respectively~(Fig.~\ref{fig_twr_scheme}).

From Equation~\eqref{eq_tdma_clockDrift}, it is feasible for Device A to correct its own local clock in order to synchronize with the local clock of device B or vice versa by adjusting the drift~(i.e., $\Delta e$ in Equation~\eqref{eq_tdma_clockDrift}) between the two local devices~(refer to the shaded areas in~Fig.~\ref{fig_twr_scheme}). Suppose the two devices have been already synchronized to each other~(i.e., they are in the same clock domain), $\Delta e$ in Equation~\eqref{eq_tdma_clockDrift} will exactly be zero. Otherwise, $\Delta e$ in Equation~\eqref{eq_tdma_clockDrift} will be either a positive or negative value, i.e., depending on the clock of one leading the other during the assessment process of the mentioned drift $\Delta e$. In terms of the three UWB localization schemes defined in Section~\ref{subsec_three_topos}, the multiplexing technique is applicable to both IGUULS topology~\cite{Vecchia2019TALLA, Tiemann2017Scalable} and BULS~\cite{Sang2019Bidirectional}. In contrast, the multiplexing is not required for the GUULS~(i.e., GNSS-like topology) because the scheme is capable of providing an unlimited number of mobile devices~(tags) for navigation purposes as already explained in Section~\ref{subsec_feat_comp}.

\subsection{Model Integration of BULS with Complementary Processes for System Enhancement} \label{subsec_complementary_setup}

Referring to Fig.~\ref{fig_system_setup_block_diag}~(b), the complementary operational processes of BULS (i.e., colored with light orange blocks in the figure) are useful mainly for the enhancement of the positioning system as mentioned in the beginning of this section~(Section~\ref{sec_system_model}). Otherwise stated, the bidirectional UWB localization system will still function properly without these complementary processes. However, the data measurement process in the real world, especially for radio-based technologies, is error-prone in nature due to several factors such as LOS, NLOS, and multi-path conditions as well as the material involved in the radio propagation path~\cite{Dardari2009Ranging}. In consequence, the degradation of location services is commonly evident in a radio-based positioning system in terms of performance. Therefore, the complementary processes addressed in this section play crucial roles in enhancing the location-oriented solutions provided by the radio-based system. Concretely, the system enhancement for BULS~(Fig.~\ref{fig_system_setup_block_diag}~(b)) can be carried out by using three methodologies in respect of modular approach: (i) the filtering process~(Section~\ref{subsec_filtering_process}), (ii) the identification process~(Section~\ref{subsec_identification}), and (iii) the mitigation process~(Section~\ref{subsec_mitigation}). Nevertheless, it should be noted that some modular processes defined in this section can be mixed or combined into a single process in practice~(Section~\ref{subsec_digest_practice_impl}).

\subsubsection{State-Space Model for a Filtering Process} \label{subsec_filtering_process}

The filtering process in the localization system including BULS~(Fig.~\ref{fig_system_setup_block_diag}~(b)) is a complementary tool for enhancing the overall performance of the system by efficiently addressing the uncertainty in raw estimation produced by a positioning algorithm. The primary reason is that no measurement data in practice are perfect, which is no exception for UWB technology. The filtering process in the system generally enables the achievement of optimized solutions for a given task in the aforementioned imperfect measurement environments. Among others, the Bayesian-based filtering techniques are especially attractive and broadly applied in radio-based location estimation systems including the UWB~\cite{Fox2003Bayesian}. Moreover, the Bayesian-based filtering process can flexibly be integrated into the multi-sensor fusion problems depending on system requirements in addition to the use-case of a typical filter in the application~\cite{Guo2022Enhanced, Feng2020Kalman}. 

\begin{figure}
    \centering
    \includegraphics[width=8.8cm]{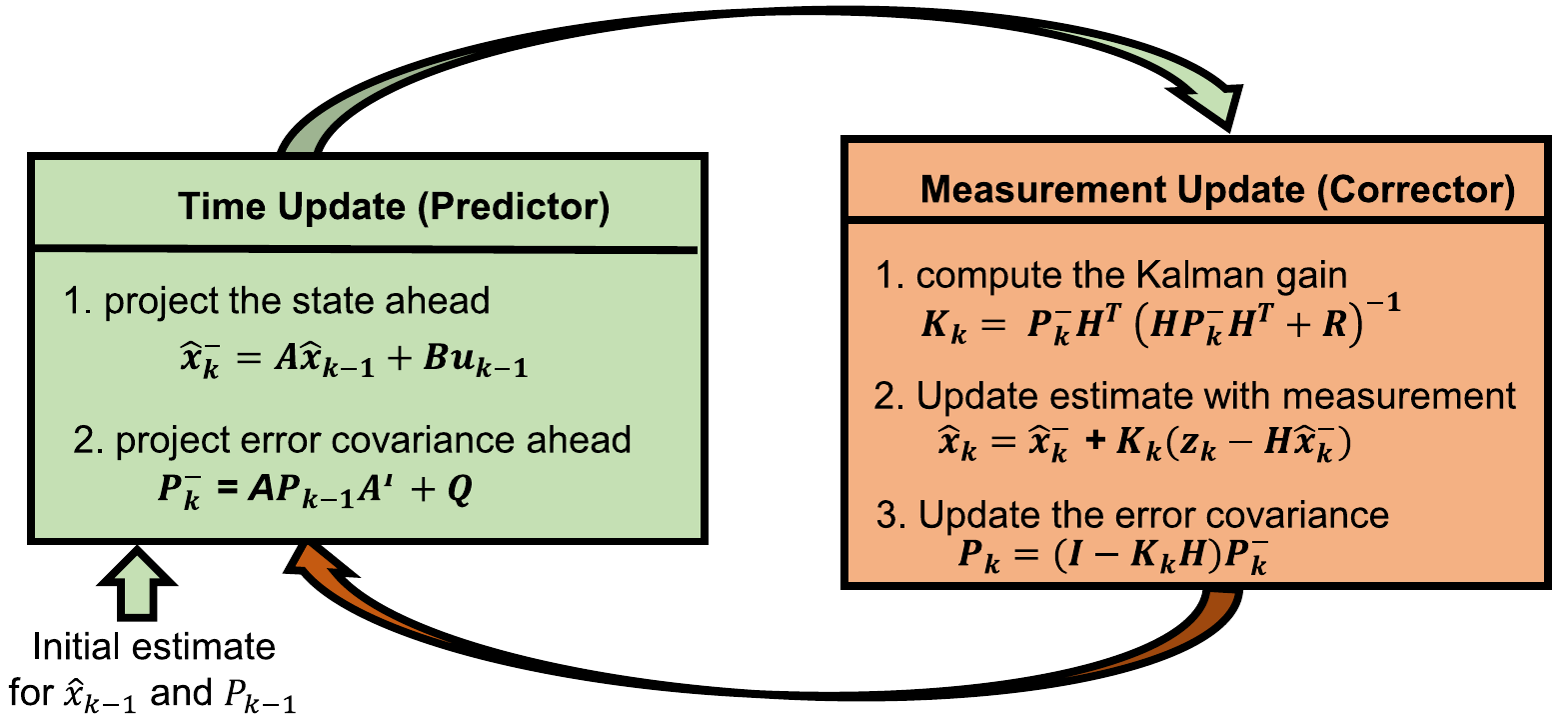}
    \vspace{-0.7cm}
    \caption{Illustration of Standard Kalman Filter Operation Process. The representation of the image was inspired by the work in~\cite{Welch2001Introduction}}
    \label{fig_filtering_kf}
    \vspace{-0.5cm}
\end{figure}

Specifically for the UWB-based positioning system, the work in~\cite{Sang2019Bidirectional, Zandian2019Ultra, Sang2019Comparative, Guo2022Enhanced, Feng2020Kalman} addressed the system implementation processes of the standard Kalman Filter~(KF) as well as its variants such as EKF and UKF, which are basically built upon the Bayesian framework. In short, the filtering process of BULS in this article is considered a Bayesian-based state-space model. Though there are other filtering methods applicable in the location estimation process such as the Butterworth filter, moving average filter, etc., Kalman-based filters are the most common and widely used one~\cite{BarShalom2004Estimation}. Fig.~\ref{fig_filtering_kf} depicted the general operational flow of standard KF~\cite{Welch2001Introduction}, which can be as a filter in the location estimation process of wireless positioning and navigation for system enhancement.

In the Bayesian-based state-space model, the dynamic nature of a system can be represented with two known mathematical models~\cite{Welch2001Introduction, RongLi2003Survey}. The first model is related to the evolution of the state of a system with respect to time and is normally termed a dynamic or system model. The second model has associated with the noisy measurement contributed to the state of the system and is commonly termed the measurement model. In general, the state and measurement models can be linear or non-linear based on the scenarios and requirements of the applications. Specifically for UWB localization, the measurement model is nonlinear due to the quadratic function in the ranging process as expressed in Equation~\eqref{eq_range_3d}. However, the state model as a linear function using Newton’s law of motion is common practice, especially in point-based wireless localization systems~\cite{Sang2022Dissertation}. The point-based system means modeling the moving targets in a location-based service by assuming point objects~(dots in the map) without any physical shapes, i.e., the estimation is normally based on the center of the object. 

Indeed, there are kinematic/motion models based on linear or nonlinear mathematical functions depending on the behavior of maneuvers~\cite{RongLi2003Survey}. Though, the most commonly used motion models for wireless positioning and navigation systems (i.e., UWB, GNSS, etc.) are the linear ones based on the constant velocity~(CV), a.k.a position velocity as well as white-noise acceleration model, or the constant acceleration~(CA), a.k.a Wiener-sequence acceleration model. In CV and CA models, it is generally assumed that the derivatives of the position in the second-order (i.e., velocity for CV) and the third-order (i.e., acceleration for CA) terms are random processes with zero mean. For brevity, we refer to the work in~\cite{Sang2022Dissertation, Sang2019Bidirectional, Sang2019Comparative, Guo2022Enhanced, Feng2020Kalman} for the detailed implementation of the said state and measurement models for KF in the UWB localization system.

\subsubsection{Identification Processes for Non-direct Path Signals in UWB} \label{subsec_identification}

The majority of errors in the UWB localization system are caused by the non-direct path signals in the ranging process~(Section~\ref{subsec_ranging_process}). This is due to the fact that the measured ranges in the UWB localization system are basically affected by the measurement conditions such as LOS, NLOS, and Multi-path~(MP). In this context, the non-direct path signals refer to the NLOS and MP conditions. In general, the propagation time of the UWB signals is slowed down when the signal needs to penetrate obstacles~(NLOS condition) and it is elongated in MP conditions~\cite{Dardari2009Ranging}. Consequently, the estimated ranges in the system basically form positive biases in the measurement compared to the actual true ones. Therefore, the identification process and mitigation process~(Section~\ref{subsec_mitigation}) of the mentioned errors due to the non-direct path signals are crucial for enhancing the system's performance. However, the identification process of ranging errors in wireless communications~(i.e., UWB localization system in particular) is very challenging and difficult to realize in practice because of the physical limit in the radio propagation channel~\cite{Dardari2009Ranging}.

In brief, the conventional techniques regarding the identification process of non-direct path signal in UWB measurement are mainly based on exploiting the statistical conditions of the received UWB signal~\cite{Sang2020Identification, Sang2022Dissertation}. The commonly used conventional techniques include statistical methods such as binary hypothesis test and kurtosis as well as exploiting the information from the received UWB signal such as signal-to-noise ratio, channel impulse response,  received signal strength, first path signal, etc.~\cite{Sang2020Identification}. Lately, machine learning techniques are regarded as attractive solutions for overcoming the mentioned fundamental limits in wireless communications~\cite{Zhang2019Deep, Zhu2020Toward}. In the context of the UWB localization system, this implies the identification and mitigation processes of the errors produced by the non-direct path signals. However, the field is generally considered in its earlier stage with open research. 

In general, the identification process of non-direct path signal in UWB localization is commonly regarded as a binary classification problem~(LOS vs. NLOS) in the literature. For instance, the identification of NLOS scenario was addressed as a binary class using the conventional technique in~\cite{Conti2014Experimental}, the classical machine learning methods in~\cite{Marano2010NLOS}, and deep learning approaches in~\cite{Park2020Improving}. On the contrary, a few works also addressed the mentioned problem as a multi-class problem~\cite{Sang2020Identification, Barral2019NLOS, Kolakowski2018Detection}, from which the research data and its corresponding source code from \cite{Sang2020Identification} were publicly provided as open access.

\begin{figure}[!t]
    \centering
    \includegraphics[width=8.8cm]{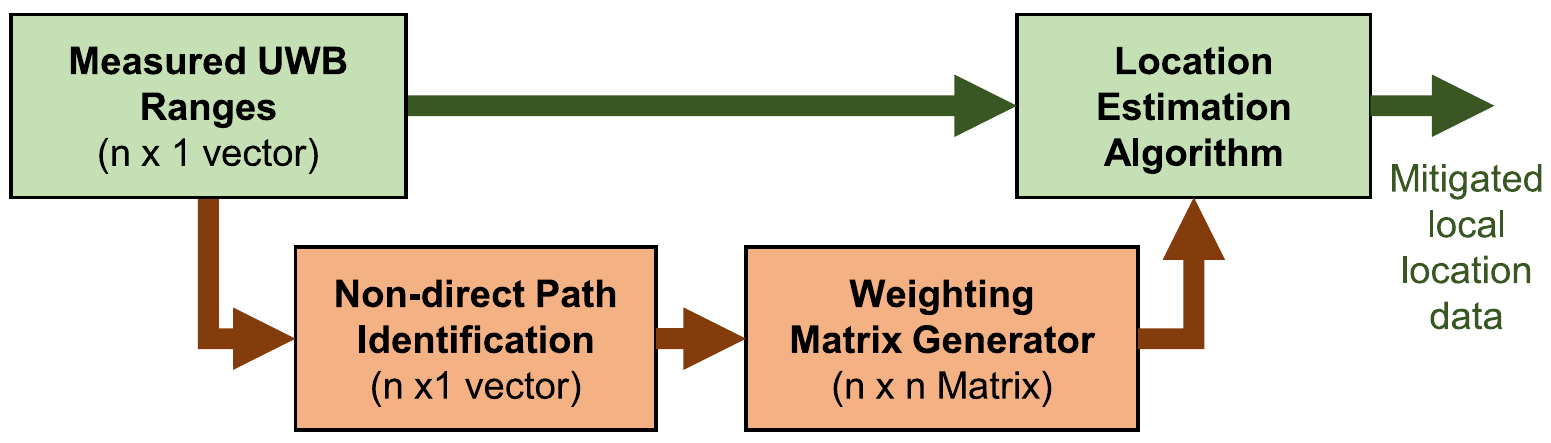}
    \vspace{-0.5cm}
    \caption{Illustration of the implementation process for a simple and effective mitigation technique for UWB localization~\cite{Sang2022Dissertation}.}
    \label{fig_mitigation_technique}
    \vspace{-0.5cm}
\end{figure}
\subsubsection{Mitigation process for non-direct path signals in UWB} \label{subsec_mitigation}

This section presents a simple, powerful, and effective technique  as well as a generic one for the mitigation of non-direct path signals in UWB localization as a modular process~(Fig.~\ref{fig_system_setup_block_diag}~(b)). In brief, the core idea of the said mitigation technique is to generate a weighting matrix based on the determined non-direct path signals in the identification process addressed in the previous section~(Section~\ref{subsec_identification}). Then, the location estimation algorithm~(e.g., Taylor-series method) makes a correction on the errors during measurements using the generated weighting matrix. The pseudo-code regarding the weighting matrix  generation process for  the mitigation of the non-direct path signals in the UWB ranging phase can be defined as in Algorithm~\ref{weighting_algorithm_multiclass}~\cite{Sang2019Comparative, Sang2022Dissertation}.  The presented algorithm shows the generation of a weighting matrix for a multi-class problem~(i.e., identification and mitigation of LOS, NLOS, and MP conditions in UWB localization). 

\begin{algorithm} 
	\begin{algorithmic}[1]
		\renewcommand{\algorithmicrequire}{\textbf{Input:}}
		\renewcommand{\algorithmicensure}{\textbf{Output:}}
		\REQUIRE \#mp \COMMENT{number of MP measurements in ranging phase} 
		\REQUIRE \#nl \COMMENT{number of NLOS measurements} 
		\REQUIRE \#r \COMMENT{total number of 
 the measurements}

 	\STATE iw $\Leftarrow $ $\dfrac{1}{\text{\#r}}$ \COMMENT{initial equal weight value, scalar}
		\STATE W $\Leftarrow$ eye(\#r) \COMMENT{initial weighting matrix}
		\IF{\#nl = 0 and \#mp=0}
		\STATE return W \COMMENT{no weighting is required}
		\ELSE
		\FOR{$i \Leftarrow$ 1 to \#r}
		\IF{ measurement $i$ is NLOS}
		\STATE W($i, i$) $\Leftarrow \dfrac{\text{iw}}{\text{2$\cdot$\#r}}$ 
        \COMMENT{relegate the initial scalar weight with a weighting factor of $\dfrac{1}{\text{2 $\cdot$ \#r}}$, which is defined for NLOS condition in this particular use-case}
		\ELSIF{ measurement $i$ is MP}
		\STATE W($i, i$) $\Leftarrow \dfrac{\text{iw}}{\text{\#r}}$ 
        \COMMENT{ here, the weighting factor for MP condition is predefined as $\dfrac{1}{\text{\#r}}$}
		\ELSE
        \STATE W($i,i$)$\Leftarrow\text{iw}+\dfrac{(\text{iw}-\dfrac{\text{iw}}{\text{$2$\#r}})\text{\#nl}+(\text{iw} - \dfrac{\text{iw}}{\text{\#r}})\text{\#mp}}{\text{\#r} - \text{\#nl} - \text{\#mp}}$
        \COMMENT{distribute the surplus from the relegated weights achieved from NLOS and MP to the LOS}
		\ENDIF
		\ENDFOR
		\ENDIF
		\RETURN W
	\end{algorithmic} 
	\caption{Pseudo-code for generating weighting matrix for mitigation of non-direct path signals in UWB~(multi-classes)}
	\label{weighting_algorithm_multiclass} 
\end{algorithm}

Firstly, the algorithm assigns equal weights to all the identifiable measured ranges~(line~$1$ in Algorithm~\ref{weighting_algorithm_multiclass}). For instance, there are three identifiable ranges namely LOS, NLOS, and MP in Algorithm~\ref{weighting_algorithm_multiclass}. The weight of a given UWB measured range is relegated from its initially given value using a predefined weighting factor if NLOS or MP conditions are identified in the measurement. The mentioned exemplary weighting factors are $1/(2\cdot\#r)$ for NLOS and $1/\#r$ for MP condition respectively, i.e., line~$8$ and $10$ in Algorithm~\ref{weighting_algorithm_multiclass}. Consequently, the surpluses~(excess) of the two relegated weights from NLOS and MP are evenly distributed to the weight of LOS conditions within the measurements~(line $12$ in Algorithm~\ref{weighting_algorithm_multiclass}). It can be shown in the algorithm that the trace of the weighting matrix~(i.e., the sum of all elements in the diagonal of the matrix) is always $1$ except for extreme cases where all measurements fall under one condition~(i.e., LOS, NLOS, or MP). In such exceptional cases, equal weights will be applied to the location estimation algorithm, which has no effect on the measurements. In addition, the mentioned weighting matrix is commonly assumed as a diagonal matrix because the measurements in the ranging process are generally independent of each other. It should be noted that the predefined weighting factors for NLOS and MP conditions can be tuned in accordance with the system requirements.

The presented mitigation technique can flexibly be applied in the vast majority of the location estimation algorithms~\cite{Sang2019Comparative, Shen2008Performance} commonly used in UWB positioning and navigation. The details of positioning methods in UWB localization as well as the implementation process of the mentioned mitigation technique on each method are beyond the scope of this article~(Section~\ref{subsec_exp_demo}). Nevertheless, interested readers are encouraged to look at the comprehensive detail of such an integration process in~\cite{Sang2022Dissertation, Yu2019Novel, Guvenc2007NLOS}. Instead, the exemplary implementation process of the said mitigation technique on the Taylor series positioning method~(Section~\ref{subsec_positioning_process}) is briefly demonstrated in this article. Fig.~\ref{fig_mitigation_technique} depicts the overview of the integration process for the mentioned mitigation technique using block diagrams. In a nutshell, the weighting matrix assigns priority levels in the measurements when location estimation is conducted, i.e., more weight to the LOS and less to the NLOS and MP conditions. As a result, the negative impacts of the non-direct path signals in the system are nicely filtered out to provide the optimized solution. For the demonstration purpose, the implementation process of the mitigation technique in the Taylor series-based location estimation algorithm can be established as in the following by modifying Equation~\eqref{matrix_form_TS}:

\setlength{\abovedisplayskip}{1pt} 
\begin{align}
	H \delta &= \Delta d     \IEEEnonumber \\
	W H \delta  &= W \Delta d    \IEEEnonumber \\
	\delta  &= (H^T W^T W H)^{-1} H^T W^T W b \IEEEnonumber  \\
	\delta  &= (H^T C H)^{-1} H^T C b 
\end{align}

where, $W$ is again the weighting matrix generated by the mitigation algorithm addressed in this section, and $C$ is $ W^T W $. Suppose that the weight produced from the inverse of the covariance matrix using the over-determined least squares~(Equation~\eqref{incremental_delta}) plays important role in the system, the optimized solution using the presented weighting matrix is still possible. In that specific case, the inverse of the covariance matrix needs to be modified in Equation~\eqref{incremental_delta}, i.e., $W^T R^{-1}W$.

\subsection{Digest on Implementation Processes in Practice} \label{subsec_digest_practice_impl}

The primary challenge for the quality assurance of radio-based location service, i.e., UWB localization in this article, is to deal with the errors produced by the non-direct path signal propagation as mentioned earlier. The main reason is that the overall system performance can drastically deteriorate due to these effects. Thus, the identification and mitigation processes of the non-direct path signals~(NLOS and MP incidents) are vital for the enhancement of location services provided by wireless positioning systems, particularly in the UWB localization system. Accordingly, there are several strategic approaches to overcome the mentioned issues in practice. 

In general, the strategy for correcting the ranging errors due to the non-direct path signals is commonly considered in two steps~\cite{Marano2010NLOS, Khodjaev2010Survey, Wu2019Neural, Yu2019Novel, Guvenc2007NLOS, Borras1998Decision, Wymeersch2012Machine}, i.e., like the modular processes addressed in the previous section. In this regard, the first step involves the identification process as in~\cite{Guvenc2007NLOS, Borras1998Decision}, and the second step is associated with the mitigation process as in~\cite{Guvenc2007NLOS, Marano2010NLOS}. 

In contrast, there exist strategic approaches that handled both the identification and mitigation processes of the mentioned non-direct path signals in one shot or a single process~\cite{Wu2019Neural, Wymeersch2012Machine, Niitsoo2018Convolutional, Jimenez2021Improving}. Basically, the mentioned one-shot approaches are commonly built upon machine learning or deep learning structures or framework. In consequence, the application area of the localization system is usually limited to a few compatible domains as the models achieved from the mentioned approach cannot easily be transferred or transformed to other areas. In other words, the results achieved from one environment using a one-shot approach cannot generally be reproduced with comparable outcomes in other environments~\cite{Sang2020Identification}. Hence, the tedious data collection process, training and validation of the machine learning models for a specific application, and re-implementation of the system are basically necessary when the localization scheme based on the one-shot approach is integrated into each unique environment. The main hurdle of this cause is due to the unavailability of the datasets yet, which are general enough to cover many real-world aspects of indoor environments or GNSS-deprived zones. Another reason is that the unforeseeable uncertainty~(i.e., the scenario unobserved in the collected training and validation data) is commonly encountered in practice during signal measurement processes in wireless communications. Whenever uncertainties are involved in the prediction of a machine learning model, caution should be taken into the integration process as the system performance can acutely deteriorate, unlike the stable world principles with well-defined rules such as Chess, Go-the board game, etc.~\cite{Gigerenzer2022How}.

Theoretically, several modular aspects~(e.g., ranging, positioning, filtering, identification, and mitigation processes, etc.) can be combined into a single process using machine learning-based one-shot approaches or traditional localization algorithms. For instance, three modular processes  of the UWB localization system namely the ranging, positioning, and filtering aspect can be considered as a single process in practice by using the Bayesian-based location estimation methods like EKF~\cite{Tiemann2016ATLAS}, UKF~\cite{Zandian2019Ultra, Sang2022Dissertation}, and Particle filter~\cite{Gonzalez2009Mobile}. Here, the inputs of the localization systems are conventionally the ranging data and the outputs are the filtered location data estimated by the applied Bayesian-based positioning algorithm.

 Alternatively, a simple select-and-proceed approach can also be established in UWB localization systems. Fundamentally, a location estimation algorithm needs at least three ranges in 2D and four ranges in 3D to estimate the position of a mobile tag as previously explained in Section~\ref{subsec_positioning_process}. However, multiple anchors are commonly used in the practical implementation process. For instance, the work in~\cite{BastidaCastillo2019Accuracy} used eight anchors in the system setup whereas Kinexon's local positioning system, i.e., an UWB-based sport analytic system currently deployed in the Handball-Bundesliga league in Germany as well as the NBA basketball league in North America, required $14$ anchors in the system setup~\cite{Blauberger2021Validation, Fleureau2020Validity}. In the mentioned systems, it is feasible to select the best candidates among the extracted available measurements in the ranging process using classic statistical techniques or machine learning-based approaches. Then, the location estimation of the interested tag is conducted with a few chosen ones by a positioning algorithm and the rest of the measurements can simply be discarded or put aside.

\section{Demonstrative Evaluation Results and SWOT Analysis} \label{sec_eval_results}

This section concisely demonstrates the comparative analysis of two system integration processes~(i.e., minimum model vs. complementary model) of BULS using the experimental data~(Section~\ref{subsec_exp_demo}). In addition, the SWOT analysis of the scheme was also discussed in Section~\ref{subsec_SWOT_analysis}.

\subsection{Experimental Demonstration of BULS based on Two Model Integration Processes} \label{subsec_exp_demo}

\begin{figure}[!t]
    \centering
    \includegraphics[width=8.8cm]{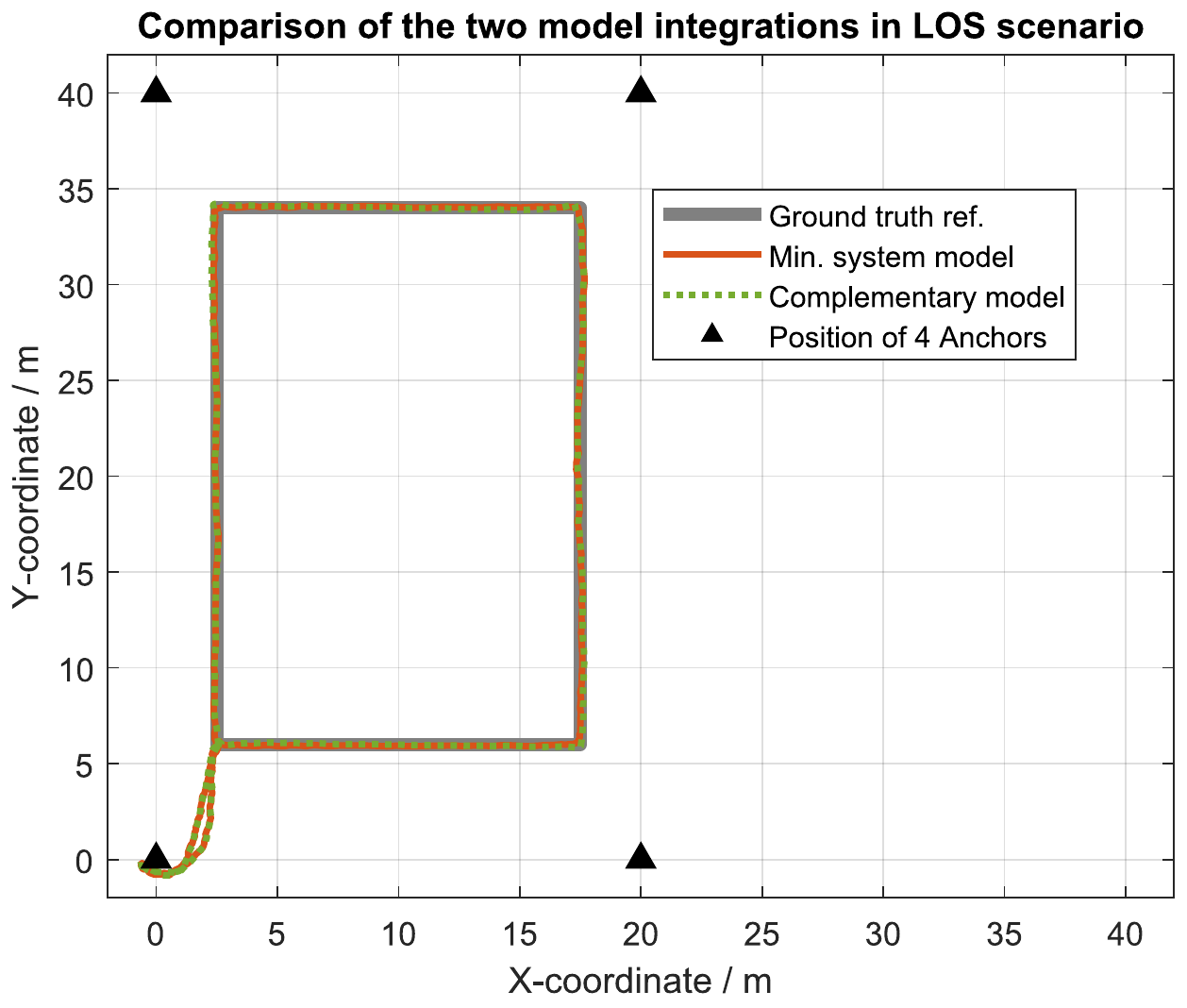}
    \vspace{-0.5cm}
    \caption{Demonstration of the two system models' integration~(minimum setup vs. complementary processes) in LOS scenario. The naming of four anchors denotes Anc. 1, 2, 3, and 4 in this figure and Fig.~\ref{fig_results_demo_NLOS}, starting from the bottom left counterclockwise.}
    \label{fig_results_demo_LOS}
    \vspace{-0.5cm}
\end{figure}

\begin{figure}[!b]
    \vspace{-0.3cm}
    \centering    \includegraphics[width=8.8cm]{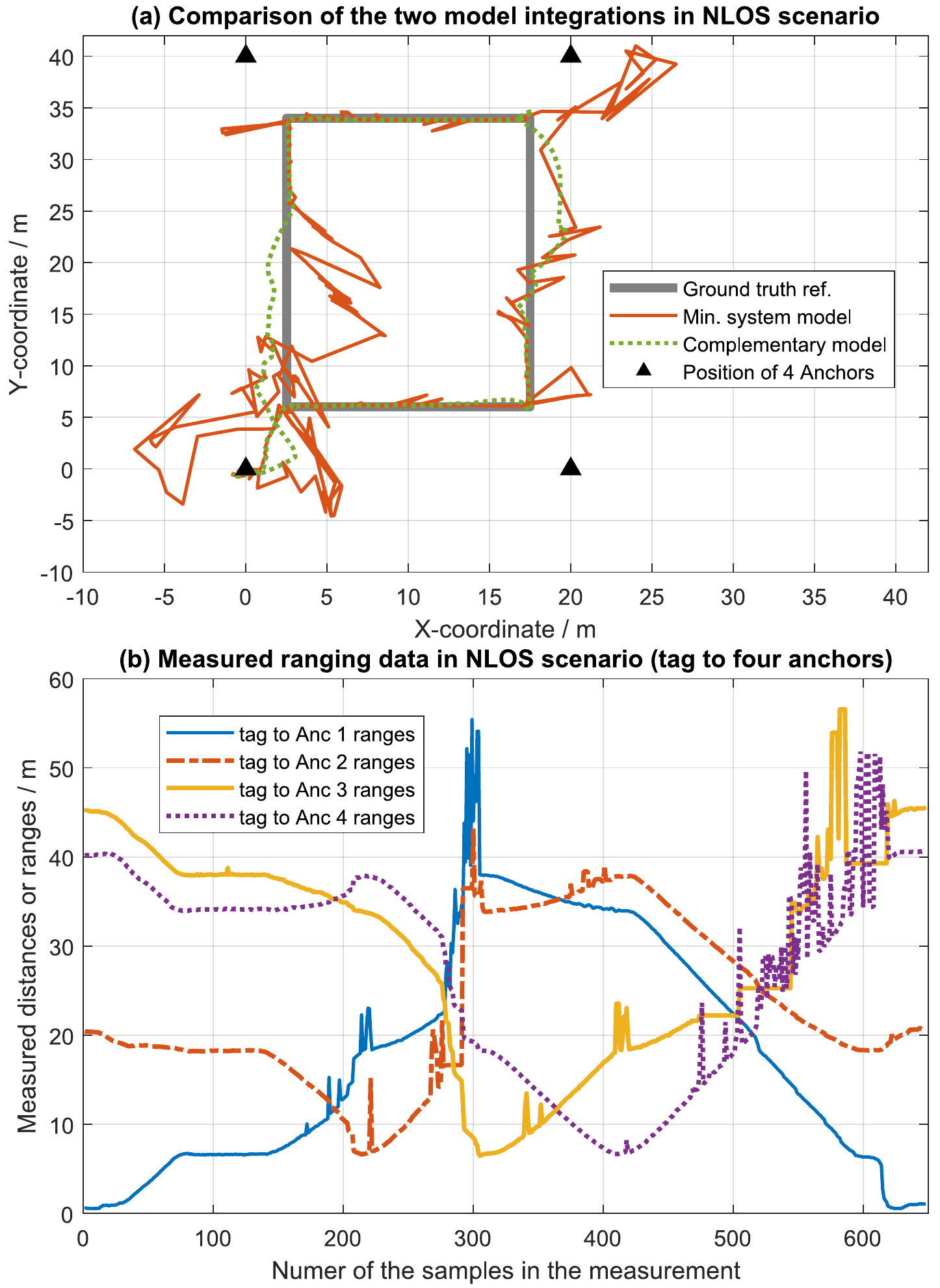}
    \vspace{-0.5cm}
    \caption{Demonstration of two system models' integration in NLOS scenario: (a) Comparison of the two system models in terms of the measured positioning data, (b) comparison of the measured four ranging data~(a tag to four anchors). The annotation of the anchors' number was counted in counter-clockwise starting from the bottom left in~(a).}
    \label{fig_results_demo_NLOS}
\end{figure}

For demonstrative analysis based on real-world measurements, we evaluated the integration of BULS based on two system models, specifically the minimum setup~(Section~\ref{subsec_minimum_setup}) versus the complementary setup~(Section~\ref{subsec_complementary_setup}), using experimental data. In this context, we deployed four anchors within an indoor sports hall for the BULS system setup. The experiment involved tracking the location of a runner (player) within the indoor sports hall, with the boundary of a basketball court (measuring 28m x 15m) serving as the ground truth reference for our evaluation. We utilized this boundary as our reference for both Line-of-Sight (LOS) and Non-Line-of-Sight (NLOS) scenarios to establish ground truth, as it allowed the player under test to easily follow these lines for our assessment. For the UWB hardware in the setup, we used the evaluation kit EVK1000 modules from Qorvo~(formerly Decawave). During the measurement process, the ranging data between the tag (i.e. the player) and four anchors was collected in two scenarios: LOS and NLOS. Then the presented location data of the player was calculated using first-order Taylor Series~(Section~\ref{subsec_positioning_process}) as a positioning technique~\cite{Sang2019Comparative, Sang2022Repo}. In the LOS scenario, the runner placed the tag device on top of his head to ensure that direct LOS UWB signals were received from the tag to all four anchors. Conversely, in the NLOS scenario, the tag device was positioned in front of the runner's chest during the data collection process. The evaluation results for the LOS scenario were presented in Fig.\ref{fig_results_demo_LOS}, while the results for the NLOS scenario are described in Fig.\ref{fig_results_demo_NLOS}.


In the LOS scenario, the evaluation results showed that there were no differences between the outcomes produced by the minimum system model and the complementary model~(Fig.~\ref{fig_results_demo_LOS}). In other words, the results revealed that the minimum and complementary models were able to precisely track the position of a runner in accordance with the trajectory of the reference~(Fig.~\ref{fig_results_demo_LOS}). Moreover, the two models gave an identical performance in terms of accurately following the trajectory of the reference in the LOS scenario. 

On the contrary, Fig.~\ref{fig_results_demo_NLOS}~(a) illustrated the comparative results of two system models integration processes~(minimum vs. complementary) based on the measured data in the NLOS scenario. Correspondingly, the measured ranging data (a tag to four anchors) concerning the NLOS scenario was also described in Fig.~\ref{fig_results_demo_NLOS}~(b). As expected, the minimum system model cannot correctly follow the trajectory provided by the reference~(Fig.~\ref{fig_results_demo_NLOS}~(a)). Moreover, the results showed that the highly unreliable and chaotic location data are evident in the minimum system model (i.e., a solid line with orange color in~Fig.~\ref{fig_results_demo_NLOS}~(a)). The cause of the mentioned chaotic nature of location data provided by the minimum system is due to the errors caused by the NLOS conditions in the ranging phase. This can be confirmed in the measured ranging data provided in Fig.~\ref{fig_results_demo_NLOS}~(b), i.e., the spikes shown in the measurement. In particular, the higher the spikes in the ranging process, the greater the error in the location estimation process. Therefore, effectively addressing the mentioned ranging error by using the identification and mitigation techniques as addressed in this article is essential for the persistent location information services in many applications. To demonstrate  this, the results achieved by the complementary model were also provided in Fig.~\ref{fig_results_demo_NLOS}~(a), i.e., the dotted line with green color. It can be stated with confidence that the location data provided by the complementary model are by far better than the minimum model. However, the results also disclosed that a perfect trajectory as in the LOS scenario cannot be achieved in the complementary model, particularly for the setup and measurements gathered in this specific scenario. This is because at least three perfect ranges in 2D are required in the location estimation processes~(Section~\ref{subsec_positioning_process}). Consequently, it is relatively hard to achieve three perfect ranges using four anchors even if the mitigation process is applied in the given experimental setup~(Fig.~\ref{fig_results_demo_NLOS}~(b)). Thus, multiple anchors in the system design are generally the preferred approach in many practical applications~(Section~\ref{subsec_digest_practice_impl}).

In summary, the primary objective of this section is to offer insights from experimental real-world measurements, particularly with regard to BULS, in a concise manner. Concerning this, we conducted an assessment of two BULS system models: the minimum setup (Section~\ref{subsec_minimum_setup}) and the complementary setup~(Section~\ref{subsec_complementary_setup}) within an indoor sports hall using the boundary of a basketball court as our reference. While both models performed well in LOS conditions~(Fig.~\ref{fig_results_demo_LOS}), the NLOS scenario revealed notable distinctions~(Fig.~\ref{fig_results_demo_NLOS}). The minimum model encountered challenges, resulting in unreliable location data due to NLOS-induced errors. In contrast, the complementary model demonstrated improved outcomes but was unable to achieve a flawless trajectory. These illustrative findings emphasize that real-world measurements using wireless communications, specifically UWB as discussed in this article, can be inherently complex and may still yield location errors, even when employing techniques like filtering, identification, and mitigation processes. This highlights the significance of employing multiple anchors, which is a common practice in real-world deployments, to mitigate ranging errors and enhance location accuracy.

\subsection{SWOT Analysis on the Bidirectional Localization Scheme} \label{subsec_SWOT_analysis}

\begin{figure}[!t]
    \centering
    \includegraphics[width=8.85cm]{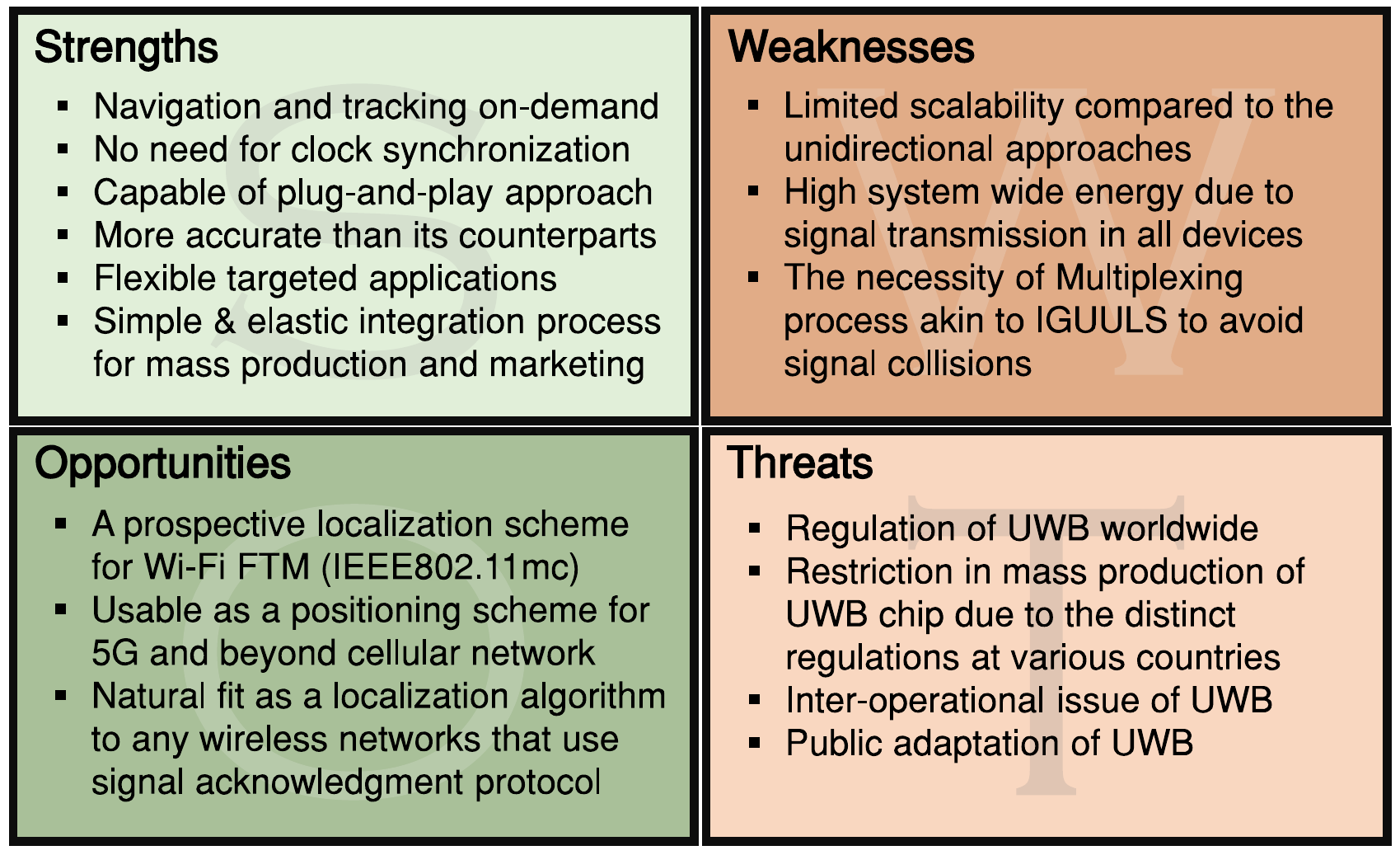}
    \vspace{-0.3cm}
    \caption{Summary of SWOT analysis for the bidirectional localization scheme in wireless positioning and navigation systems.}
    \label{fig_SWOT}
\end{figure}

SWOT~(Strengths, weaknesses, opportunities, and threats) analysis is a tool commonly used for evaluating technology, business, or the performance of an organization. The goal of SWOT analysis is to identify the inhibitors and enhancers of the organization or technology based on internal and external factors~\cite{Leigh2009SWOT}. In this regard, strengths are enhancers of internal factors~(i.e., insiders' impact) whereas weaknesses are the inhibitors. By contrast, opportunities are enhancers of external factors~(i.e., outsiders' impact) whereas threats are inhibitors. In this article, we applied the SWOT analysis for 
 the assessment of the bidirectional localization scheme. Fig.~\ref{fig_SWOT} describes the summary of the SWOT analysis regarding BULS for wireless positioning and navigation. Though SWOT analysis is not a quantitative methodology, it is very helpful for assessing the actual standing of technology or organization~(here, it is a scheme) and for revealing its potential as well as the areas where effective resolve for a solution is necessary~\cite{Alarifi2016Ultra}.

\subsubsection{Strengths}

The core advantage of BULS is its ability to flexibly and/or elastically combine two unique perspectives of wireless localization~(i.e., navigation vs. tracking) using a single localization scheme~(Section~\ref{subsec_feat_comp}). This enables the location information to be extracted in both the tag~(a mobile object) and anchors~(a fixed infrastructure). Moreover, the clock synchronization between UWB transceivers in the system setup, i.e., the main hurdle in GUULS and IGUULS topologies, is not necessary for BULS. This effectively eliminates the resources and sophisticated system setup required for the synchronization process. Thus, plug-and-play system integration can be established in BULS akin to the work presented in~\cite{Chen2022PnPLoc}. In addition, the location outcomes provided by the BULS are generally more accurate than its counterpart unidirectional approaches~(GUULS and IGUULS) because synchronization errors cannot be completely eliminated in practice~\cite{Vecchia2019TALLA, Tiemann2019ATLAS}. In accordance with the mentioned strengths and the elastic nature of the scheme, BULS opens endless potential applications, especially in diversified GNSS-denied environments. Moreover, the scheme can also be seen as a paradigm shift in the way location services are observed as mentioned several times in this article. Therefore, the scheme is suitable for general-purpose precise location services targeted for both navigation and tracking in one shot. The mentioned feature is very attractive for mass production and marketing strategy of location-aware products. 

\subsubsection{Weaknesses}

Despite the several advantages of BULS, the scheme does have several weaknesses as well. The most obvious one is the scalability of the scheme in terms of available tags and anchors within the system compared to its rival unidirectional topologies namely GUULS and IGUULS (Section~\ref{subsec_feat_comp}). Moreover, the system-wide energy consumption of the scheme is also relatively higher than its opponents~(GUULS and IGUULS) as already expressed in Section~\ref{subsec_feat_comp}. Another shortcoming of the scheme is that BULS, i.e., IGUULS as well, needs a multiplexing scheme in order to avoid possible signal collisions~(Section~\ref{subsec_multiplexing_process}). This leads to the limitation in terms of the volume of tags observable in the system for BULS and IGUULS. In contrast, GUULS is empowered by a simple system architecture without the need to use a multiplexing scheme as well as exceptional scalability in terms of observable tags in the system~(i.e., there is no limitation for the number of tags in this scheme).

\subsubsection{Opportunities}

The bidirectional localization scheme retains several potentials to be used as a prospective positioning topology for location-oriented services in many radio-based technologies. A good example is the WiFi FTM defined by IEEE 802.11mc standard, where TWR is intended to enforce in the ranging process as mentioned earlier in Section~\ref{sec_potential_beyond_uwb}. Moreover, the bidirectional localization scheme offers several demanding characteristics for realizing precise and customizable location-aware services in 5G and beyond wireless networks. Besides, the present allowable spectrum usage of UWB technology~(Fig.~\ref{fig_uwb_regulation}) generally collides with the spectrum of interest in 5G/6G wireless networks, which is still under active research and discussion to date~(Section~\ref{sec_potential_beyond_uwb}). Therefore, the bidirectional localization scheme can play a significant role in many applications in 5G/6G-based precise positioning and navigation systems. In addition, the bidirectional localization scheme naturally fits as a positioning system for any wireless networks that use the signal acknowledgment protocol in the data communication process. Here, the message related to the acknowledgment can generally be made use of as a TWR procedure. Hypothetically, nearly all protocols in wireless networks use a signal acknowledgment method for secure and reliable communication flows between network devices.

\subsubsection{Threats}

The major challenge that prevents the boom of UWB technology~(i.e., regardless of  the topology) rapidly in consumer applications may include the distinct regulations defined worldwide. For instance, the frequency spectrum of the UWB that permits the maximum allowable EIRP worldwide~(i.e., \SI{-41.3}{\dBm} except in China, where it is \SI{-41.0}{\dBm}) is only between \SI{7.25}{\GHz} and \SI{8.5}{\GHz} as described before in Section~\ref{sec_potential_beyond_uwb} and in Fig.~\ref{fig_uwb_regulation}. The rest of the spectrum is sharply diversified based on the country or region~(see the detail about the regulation for other countries or regions in~\cite{ETSI2019Short}). The mentioned distinct restriction of UWB spectrum usage in various countries is, indeed, a stumbling block to the mass production of UWB chips for the manufacturer as well as for the service providers. Moreover, the interoperation or co-existence of UWB with narrowband technologies is commonly separated by restricting the allowable EIRP of the UWB. This may raise a significant interoperation issue in cases like the use of a device that was bought from a different country or region. The concern is more intense for the integration of UWB chips on everyday consumer electronics like Smartphones, which are normally carried as belongings during traveling or business trips. This leads to the next common threat generally encountered in the major shift in technologies, i.e. the public adaption of the technology~(i.e., UWB in this article) as day-to-day consumable electronic products.

\section{Related Work} \label{sec_related_work}

As previously discussed, UWB technology has become a popular choice for precise location estimation in GNSS-deprived positioning and navigation systems mainly due to its high-accuracy localization capability, i.e., typically at around \SIrange{10}{20}{\cm}~\cite{Sang2019Numerical, Sang2019Comparative}. Many surveys and reviews in the literature have explored various aspects of UWB technology, including its theories, techniques, the technology itself, its applications, and its future perspective~\cite{Gezici2009Position, Xu2006Position, Ridolfi2021Self, Mazhar2017Precise, Alarifi2016Ultra, Shi2016Survey, Elsanhoury2022Precision}. However, those surveys and reviews tend to address a wide range of topics, attempting to cover various methodologies and techniques, often presenting them in brief paragraphs. As a result, they often overlook in-depth studies, especially concerning less common topologies and techniques, as well as those that are considered conventional. For instance, the surveys in~\cite{Ridolfi2021Self, Alarifi2016Ultra, Shi2016Survey} and the review in~\cite{Mazhar2017Precise, Elsanhoury2022Precision} provide a broad overview of methodological, algorithmic, and implementation aspects of UWB technology. However, they lack detailed insights into each presented method, which may leave interested readers in need of further extensive study.

Furthermore, literature reviews are typically conducted within a defined topic. In the study of~\cite{Mazhar2017Precise}, a comprehensive review of UWB technology was performed, encompassing all radio-based technologies in IPS. Conversely, the authors in~\cite{Elsanhoury2022Precision} conducted a specific literature review on UWB-based precise positioning, focusing on its application in smart logistics, by providing a comprehensive survey of academic literature in this field. Similarly, the study in~\cite{Ridolfi2021Self} focuses on a literature survey of UWB localization, particularly for the automatic positioning system of anchor nodes, also referred to as self-calibration or self-localization of UWB anchors. The primary goal of automatically positioning these anchors in UWB technology is to reduce deployment costs where manual measurement and tuning processes are labor-intensive and error-prone~\cite{Ridolfi2021Self, Corbalan2023SelfLocalzation}. A recent study~\cite{Corbalan2023SelfLocalzation} delves into anchor self-localization for multi-hop scenarios, as opposed to the more common single-hop scenarios where UWB signals from tags can reach all anchors, or vice versa. Differing from the approaches mentioned earlier, this article offers a comprehensive review of a specific UWB topology or localization scheme. This scheme is often considered a conventional method, yet its impact and implications are not adequately described in many papers in the literature.

As previously discussed in Section~\ref{sec_SOTA_UWB_topos}, time-based localization schemes are widely utilized in practical UWB deployment, with a particular focus on measurement techniques based on ToA and TDoA. In recent years, there has been a vast exploration of time-based UWB localization strategy related to the GNSS-like schemes denoted as GUULS in this article~(aka DL-TDoA)~\cite{Ledergerber2015Robot, Zandian2017Robot, Santoro2023UWB, Paetru22023FlexTDOA, Yang2022VULoc, Elsanhoury2022Precision, Hamer2018Self, Corbalan2019Chorus, Grosswindhager2019SnapLoc, Santoro2021Scale} and the inverted GNSS-like scheme denoted as IGUULS~(aka UL-TDoA)~\cite{Zwirello2012UWB, Tiemann2016ATLAS, Tiemann2017Scalable, Tiemann2019ATLAS, Vecchia2019TALLA, Friedrich2021Accurate}. However, BULS is commonly referenced in the literature~~\cite{Tiemann2016ATLAS, Zandian2017Robot, Hamer2018Self, Ledergerber2015Robot, Zwirello2012UWB} as a conventional method, often without providing detailed information on its specific implementation and implications, assuming that readers are already familiar with the concept. To the best of the authors' knowledge, a comprehensive explanation of BULS, from its theoretical foundations to practical use cases, is currently lacking in the literature. Furthermore, BULS has not traditionally been considered a strategic localization scheme, which has resulted in its unique properties and potential applications being under-explored. Therefore, this article aims to provide a thorough review of BULS as an elastic positioning topology for GNSS-deprived zones without neglecting the importance of its counterparts. The review covers the fundamental concepts, starting from the basics (Section~\ref{subsec_minimum_setup} and \ref{subsec_complementary_setup}), and extends to its practical applications (Section~\ref{subsec_digest_practice_impl}) using a modular design principle approach. Moreover, we aim to make the content easily understandable and accessible to a broad audience interested in UWB, requiring minimal prior knowledge.

\section{Concluding Remarks} \label{sec_conclusion}

UWB has been considered one of the most promising technologies for precise positioning systems in GNSS-deprived indoor environments. Nowadays, miniature UWB chips are distributed at relatively low prices in the electronic markets by the leading manufacturers in the field including Decawave~(i.e., now Qorvo), NXP, Bespoon, Ubisense, etc. Recently, the UWB chip has been integrated into the smartphones pioneered by Apple and followed by other vendors. This signifies that the boom of UWB technology will soon be evident in consumer electronic markets in our daily lives. Therefore, it is crucial to address the UWB-based localization schemes from different aspects 
in order to meet the diverse requirements of the applications in GNSS-deprived zones.

In this article, we comprehensively review the system model of an elastic positioning scheme for UWB-based localization systems that we named bidirectional UWB localization. In principle, the bidirectional UWB localization scheme is one of the three design integration processes for a time-based UWB positioning system as thoroughly discussed in Section~\ref{sec_SOTA_UWB_topos}. However, the bidirectional UWB scheme was largely overlooked as a strategic localization topology in the literature. In consequence, the potential of the scheme for precise location-aware services in practice was excessively overshadowed. The primary key benefit of the bidirectional UWB localization scheme is that it can be used for both navigation and tracking tasks on demand in the application. This gives the scheme a powerful ability to flexibly or elastically combine two unique positioning perspectives~(i.e., navigation and tracking) within just a single scheme. In fact, the ability to combine the mentioned two unique natures into a single scheme is a paradigm shift in the way location-aware services are commonly observed. Consequently, the bidirectional localization scheme enables many unique and new prospective applications in the diverse fields of GNSS-deprived environments by banding the navigation and tracking system as one common process. 

Moreover, the article also highlights the prospect of the bidirectional scheme in location services beyond UWB technology such as the positioning method for the WiFi fine-time measurement, the 5G/6G wireless mobile network, etc. Besides, the bidirectional localization scheme described in this article has the potential to enable new applications, products, and systems in location-aware services, which are unforeseeable and unaddressed in GNSS-deprived indoor environments yet.


\section*{Acknowledgment}
This work was supported in part by the Cluster of Excellence Cognitive Interaction Technology `CITEC'(EXC 277) at Bielefeld University, funded by the German Research Foundation~(DFG) and in part by the VEDLIoT project, which received funding from the European Union’s Horizon 2020 research and innovation programme under grant agreement number 957197. The work of Cung Lian Sang was also partly supported by the German Academic Exchange Service~(DAAD). The authors are responsible for the contents of this publication. 

We also sincerely thank the anonymous reviewers for their valuable suggestions and feedback to improve the manuscript.

\section*{References}

    


\begin{IEEEbiography}
[{\includegraphics[width=1in,height=1.25in,clip,keepaspectratio]{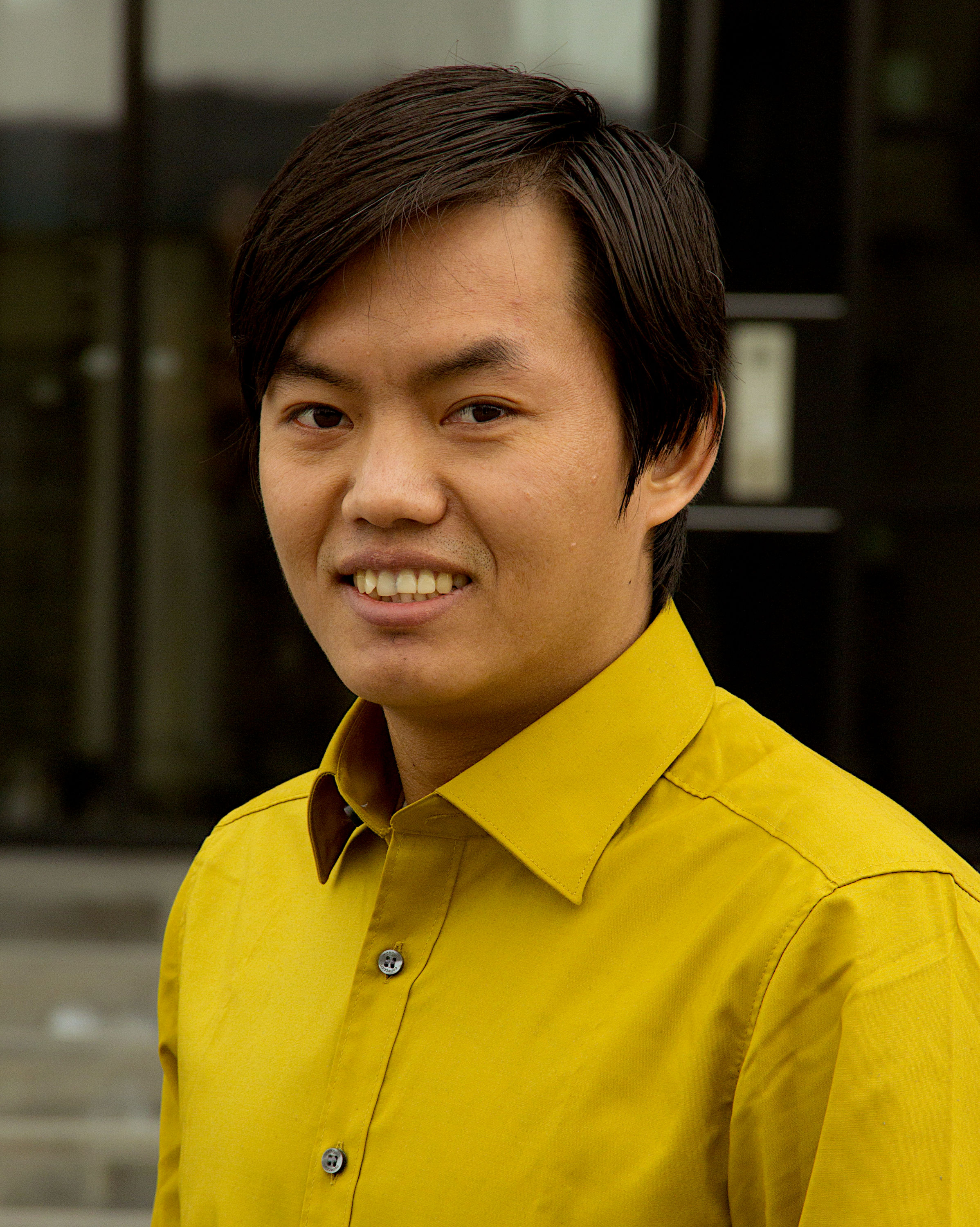}}]{Cung Lian Sang} received his B.E. and M.E. degrees in electronic engineering from Mandalay Technological University (MTU), Mandalay, Myanmar in 2006 and 2008 respectively. In 2012, he obtained his M.Sc. degree in radar and radio technology from National Research University, Moscow, Russian Federation. He acquired his Dr.-Ing. degree in electrical engineering on precise indoor localization systems using UWB technology from Bielefeld University, Bielefeld, Germany, in 2022. He is currently a postdoctoral researcher at the Congitronics and Sensor Systems group of CITEC. His research interests include wireless communications, UWB, indoor localization, machine learning, and tactile internet. 
\end{IEEEbiography}

\begin{IEEEbiography}
[{\includegraphics[width=1in,height=1.25in,clip,keepaspectratio]{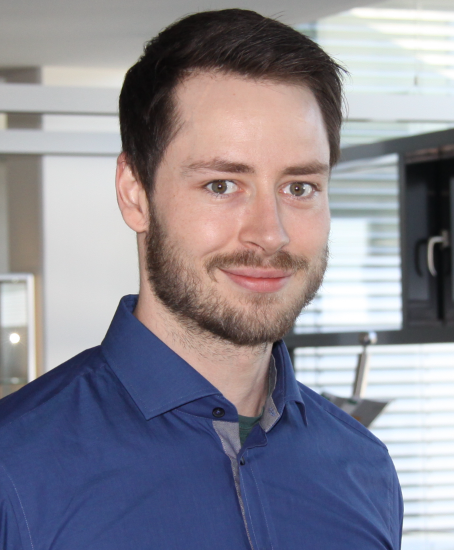}}]{Michael Adams}  earned his B.Sc. and M.Sc. degrees in computer science from Bielefeld University, Germany in 2012 and 2015, respectively.He acquired his Ph.D. in computer science at the Cognitronics and Sensor Systems group within CITEC in 2023. His primary research areas encompass machine vision and wireless localization, displaying a keen interest in the interplay of technology and spatial understanding. Additionally, Michael is deeply engaged in machine learning, applying these principles to sports data science, where he explores the potential for data-led decision-making and performance prediction.
\end{IEEEbiography}

\begin{IEEEbiography}
[{\includegraphics[width=1in,height=1.25in, clip,keepaspectratio]{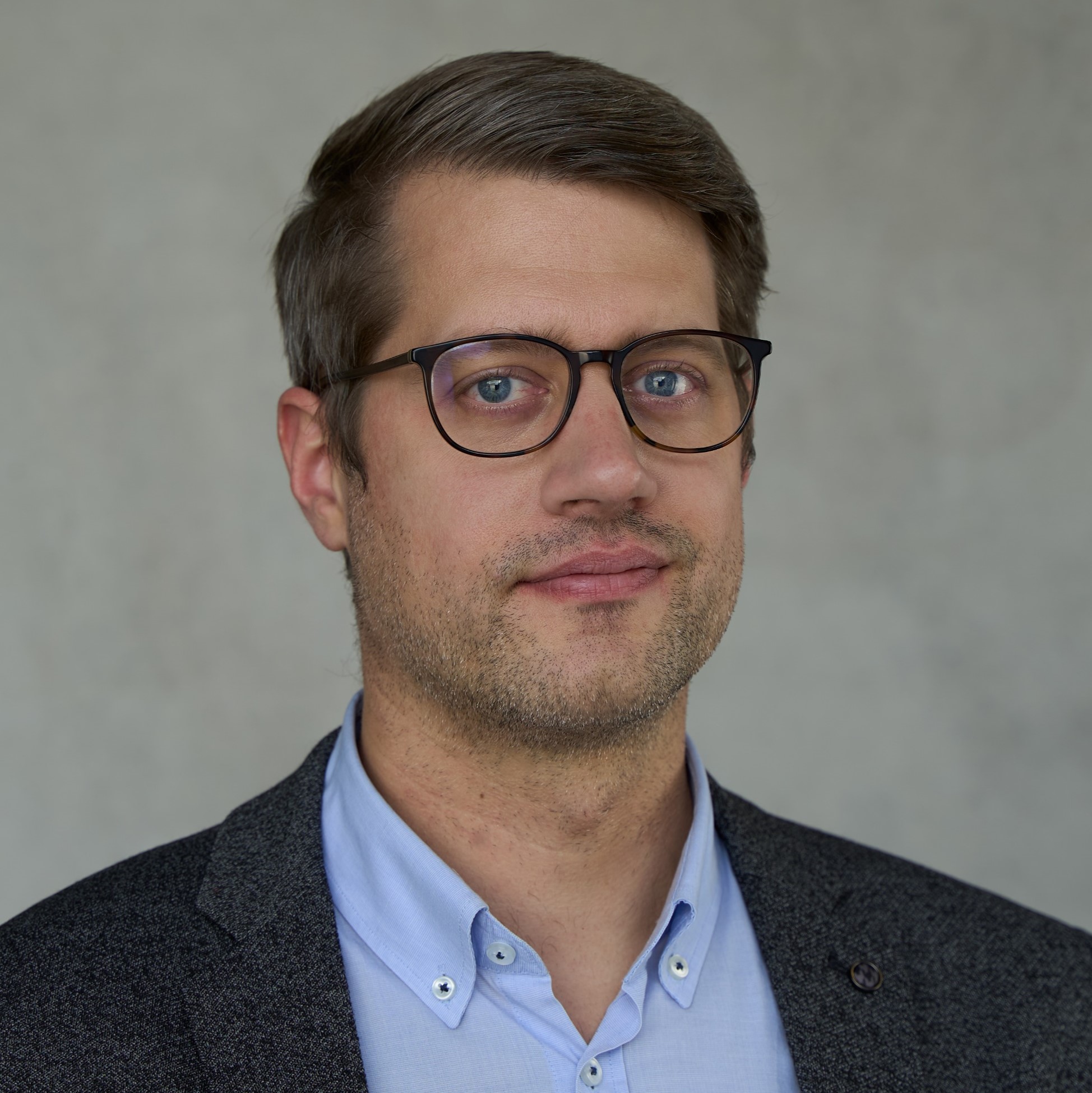}}]{Marc Hesse} (Member, IEEE), received his Diploma degree in electrical engineering from the University of Paderborn, Germany, in 2010. He acquired his Dr.-Ing. degree in electrical engineering in model-based design methodology for wireless sensor nodes from Bielefeld University, Bielefeld, Germany, in 2021. He is currently a PostDoc at the Cognitronics and Sensor Systems group of the CITEC. His research interests lie in the systematic and model-based development of embedded systems.  
\end{IEEEbiography}

\begin{IEEEbiography}
[{\includegraphics[width=1in,height=1.25in, clip,keepaspectratio]{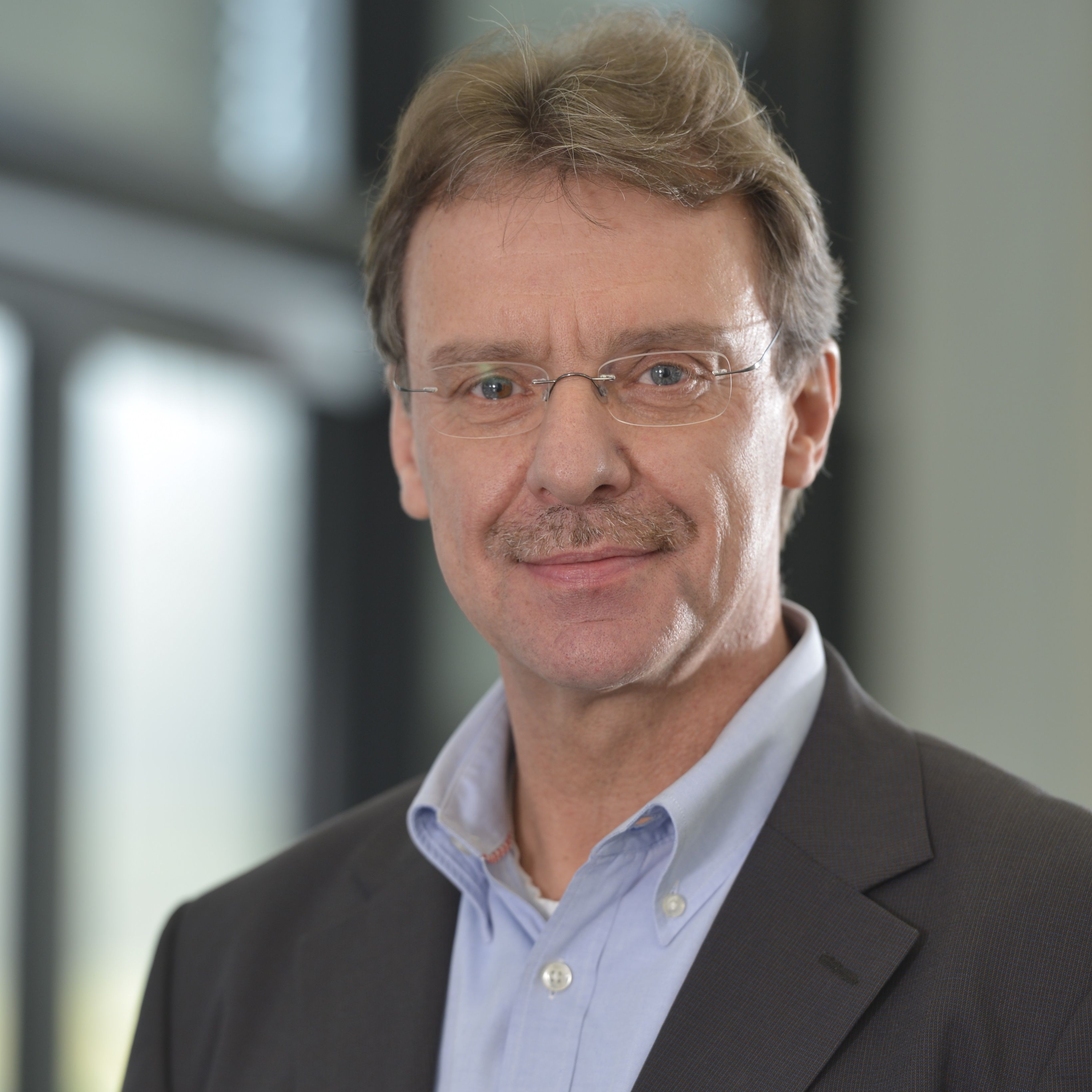}}]{Ulrich R\"uckert} (Member, IEEE),  received the Diploma degree in computer science and the Dr.-Ing. degree in electrical engineering from the University of Dortmund,
Germany, in 1984 and 1989, respectively. From 1985 to 1994 he worked in the Faculty of Electrical Engineering, University of Dortmund,
and with the Technical University of Hamburg-Harburg, Germany. In 1995 he joined as a full professor the Heinz Nixdorf Institute, University
of Paderborn, Germany, heading the research group System and Circuit Technology and working on massive-parallel and resource-efficient systems. Since 2009 he is a professor with Bielefeld University, Germany heading the Cognitronics and Sensor Systems group of the CITEC. His main research interests include now bio-inspired architectures for nanotechnologies and cognitive robotics.
\end{IEEEbiography}

\end{document}